\documentclass[usenatbib]{mn2e}
\usepackage{graphicx}
\usepackage{ifthen}
\usepackage{url}

\def\ltsima{$\; \buildrel < \over \sim \;$}
\def\lta{\lower.5ex\hbox{\ltsima}}
\def\gtsima{$\; \buildrel > \over \sim \;$}
\def\simgt{\lower.5ex\hbox{\gtsima}}
\def\kms{{\rm\,km\,s^{-1}}}

\def\kpc{{\rm\,kpc}}

\def\msun{{\rm\,M_\odot}}
\def\lsun{{\rm\,L_\odot}}

\def\pc{{\rm\,pc}}

\newcommand{\fmmm}[1]{\mbox{$#1$}}
\newcommand{\scnd}{\mbox{\fmmm{''}\hskip-0.3em .}}

\newcommand{\mcnd}{\mbox{\fmmm{'}\hskip-0.3em .}}
\def\Aa{\; \buildrel \circ \over {\rm A}}
\def\AA{$\; \buildrel \circ \over {\rm A}$}

\def\deg{^\circ}

\def\s{\ifmmode \widetilde \else \~\fi}
\def\={\overline}

\def\spose#1{\hbox to 0pt{#1\hss}}

\def\lta{\mathrel{\spose{\lower 3pt\hbox{$\mathchar"218$}}
     \raise 2.0pt\hbox{$\mathchar"13C$}}}
\def\gta{\mathrel{\spose{\lower 3pt\hbox{$\mathchar"218$}}
     \raise 2.0pt\hbox{$\mathchar"13E$}}}
\def\Dt{\spose{\raise 1.5ex\hbox{\hskip3pt$\mathchar"201$}}}    
\def\dt{\spose{\raise 1.0ex\hbox{\hskip2pt$\mathchar"201$}}}    

\def\dotsfill{\leaders\hbox to 1em{\hss.\hss}\hfill}

\def\FeH{{\rm[Fe/H]}}

\loadboldmathitalic 
\title[A spectroscopic survey of faint Galactic satellites]{A Keck/DEIMOS spectroscopic survey of faint Galactic
satellites: searching for the least massive dwarf galaxies}
\author[N. F. Martin et al.] {N. F. Martin$^{2}$, R. A. Ibata$^{3}$, S. C. Chapman$^{4,5,6}$, M. Irwin$^{4}$ \& G. F.
Lewis$^{7}$ \\
$^{2}$ Max-Planck-Institut f\"ur Astronomy, K\"onigstuhl 17, D-69117, Heidelberg, Germany\\
$^{3}$ Observatoire de Strasbourg, 11, rue de l'Universit\'e, F-67000, Strasbourg, France\\
$^{4}$ Institute of Astronomy, Madingley Road, Cambridge, CB3 0HA, U.K.\\
$^{5}$ Department of Physics and Astronomy, University of Victoria, Victoria, B.C., V8P 1A1, Canada\\
$^{6}$ Canadian Space Agency Fellow\\
$^{7}$ Institute of Astronomy, School of Physics, A29, University of Sydney, NSW 2006, Australia\\
}

\date{\today}
\begin{document} 
\maketitle 
\begin{abstract} 
We present the results of a spectroscopic survey of the recently discovered faint Milky Way satellites Bo\"otes,
Ursa Major~I, Ursa Major~II and Willman~1. Using the DEep Imaging Multi-Object Spectrograph mounted on
the Keck~II telescope, we have obtained samples that contain from $\sim15$ to $\sim85$ probable members of these
satellites for which we derive radial velocities precise to a few $\kms$ down to $i\sim21-22$. About half of these stars
are observed with a high enough S/N to estimate their metallicity to within $\pm0.2$~dex. The characteristics of all the
observed stars are made available, along with those of the Canes Venatici~I dwarf galaxy that have been analyzed
in a companion paper.\\
From this dataset, we show that Ursa Major~II is the only object that does not show a clear radial velocity peak.
However, the measured systemic radial velocity ($v_r=115\pm5\kms$) is in good agreement with recent simulations in which
this object is the progenitor of the recently discovered Orphan Stream. The three other satellites show velocity
dispersions that make them highly dark-matter dominated systems (under the usual assumptions of symmetry and virial
equilibrium). In particular, we show that despite its small size and faintness, the Willman~1 object is not a globular
cluster given its metallicity scatter over $-2.0\lta\FeH\lta-1.0$ and is therefore almost certainly a dwarf galaxy or
dwarf galaxy remnant. We measure a radial velocity dispersion of only $4.3_{-1.3}^{+2.3}\kms$ around a systemic velocity
of $-12.3\pm2.3\kms$ which implies a mass-to-light ratio of $\sim700$ and a total mass of $\sim5\times10^5\msun$ for
this satellite, making it the least massive satellite galaxy known to date. Such a low mass could mean that the
$10^7\msun$ limit that had until now never been crossed for Milky Way and Andromeda satellite galaxies may only be an
observational limit and that fainter, less massive systems exist within the Local Group. However, more modeling and an
extended search for potential extra-tidal stars are required to rule out the possibility that these systems have not
been significantly heated by tidal interaction.
\end{abstract}
 

\section{Introduction}
\footnotetext[1]{The data presented herein were obtained at the W.M. Keck Observatory, which is operated as a scientific
partnership among the California Institute of Technology, the University of California and the National Aeronautics and
Space Administration. The Observatory was made possible by the generous financial support of the W.M. Keck Foundation.}

The discrepancy between observed satellite galaxies in the Local Group and the number of dark matter halos
that are produced in simulations of such groups of galaxies is a well-known problem of the currently preferred
$\Lambda$CDM cosmology \citep{klypin99,moore99}. Various explanations have been put forward to explain this difference
of one to two orders of magnitudes between the observed and simulated number of satellites, with most of them assuming
that the satellite dwarf galaxies observed within the Local Group are surrounded by massive dark matter halos and that
they become even more highly dark-matter dominated as they become fainter \citep[e.g][]{bullock00,stoehr02}. In this
way, even faint satellites would reside in massive dark matter halos and would only represent the more massive end
of the distribution of dark halos found in simulations, hence reconciling observations and simulations.

The central velocity dispersion of spherical systems in virial equilibrium can be used to derive the total mass of the
system \citep[e.g.][]{illingworth76,richstone86}. Even though dwarf galaxies orbiting within the Local Group may
not be in virial equilibrium or perfectly spherical systems, the central velocity dispersion has been shown to be a
good indicator of the instantaneous mass of the galaxy \citep[e.g.][]{oh95,piatek95}. Hence spectroscopic observations
of faint dwarf galaxies discovered these past few years within wide field surveys, in particular with the
Sloan Digital Sky Survey (SDSS), have been rapidly conducted. They seem to show a mass limit of $\sim10^7\msun$
under which no dwarf galaxy can be found, meaning these satellites would indeed inhabit massive dark matter halos. In
particular, one can cite masses of $\sim10^7$ for Ursa Major~I \citep{kleyna05}, $\sim2\times10^7$ for Andromeda~IX
\citep{chapman05} and $\sim1\times10^7$ for Bo\"otes \citep{munoz06} in solar units. The existence of such a mass limit
would confirm the empirical relation defined by \citet{mateo98} for satellites that are brighter than Draco ($M_V=-8.8$)
and Ursa Minor ($M_V=-8.9$) in the Local Group: $M/L=2.5+ 10^7/(L/\lsun)$ where $M/L$ is the mass-to-light ratio of the
galaxies and $L$ their luminosity. Some groups have been working on the theoretical aspects of the existence of such a
limit, in order to explain it in the $\Lambda$CDM context. In particular, the recent work of \citet{read06} explains this
limit by the effect of supernovae winds that produce a sharp drop of $\sim2$ orders of magnitude of the stellar mass
over the total (stellar and dark matter) mass range $3-10\times10^7\msun$. They conclude their work by predicting that
galaxies a few orders of magnitude fainter than Draco, Ursa Minor or even Ursa Major~I ($M_V\sim-6.75$;
\citealt{willman05b}) should still have total masses within the $2-10\times10^7\msun$ range.

However, the measured mass distribution of faint dwarf galaxies within the Local Group has until now been mainly limited
by the low number of such objects for which a radial velocity survey has been conducted. With the recent discovery of
$\sim15$ new satellites around the Milky Way \citep{willman05a,willman05b,belokurov06a,belokurov06c,zucker06b,zucker06c}
and the Andromeda galaxy \citep{zucker04,chapman05,martin06,zucker06a}, all of them with $M_V\gta-8.0$, it is now
possible to populate the faint end of the satellite mass distribution and see if the $10^7\msun$ mass limit still holds
for such faint systems and if they are indeed highly dark matter dominated. As a first step toward this end, we have
used the DEep Imaging Multi-Object Spectrograph (DEIMOS) on Keck~II to derive radial velocities and metallicities for
stars within the Bo\"otes (Boo), Canes Venatici~I (CVnI), Ursa Major~I (UMaI), Ursa Major~II (UMaII) and Willman~1 (Wil1)
Milky Way satellites. The analysis of our observations is performed here for all these objects, except for CVnI which has
been analyzed in a companion paper \citep{ibata06}, where we presented in more detail its peculiar kinematic behaviour.
The outline of the paper is as follows: \S~2 presents the observing strategy, the observations and how they were reduced
; \S~3, \S~4, \S~5, \S~6 and \S~7 are dedicated to the analysis of Boo, CVnI, UMaI, UMaII and Wil1 and we conclude in
\S~8.

\section{Observations}

Target stars were selected for observation from the point sources in the fourth Data Release (DR4) of the SDSS (see Fig.~1).
Stars of highest priority were chosen within a polygon defined by eye around the probable red giant branches of the
target satellites (see Fig.~2), while all other available point-sources within the fields were assigned a lower
priority. The DEIMOS configuration program was then used to automatically design the masks.

For Boo, CVnI and UMaI, two DEIMOS masks were observed, slightly offset in declination, whereas only one field in each
of UMaII and Wil1 were observed. A planned Northern mask in UMaII, was not observed due to time constraints, which
accounts for the offset in Figure~1. Each mask was observed with 3 exposures of 1200\,s with the 1200 lines/mm grating
during the nights of 27-28 May 2006. These settings give access to the 650-900\,nm spectral region and along with 
slits of 0.7 arcsec width result in 1\AA\ resolution spectra. Contrary to the normal operations with DEIMOS, we
observed an arc-lamp spectrum immediately before and after each set of science frames, in order to ensure a better
wavelength calibration than is possible from day-time arc-lamp exposures.

The science spectra were extracted and processed in an identical manner to that described in \citet{ibata06}. In
particular, the final stage comprised a Gaussian-fit to each of the Ca~II triplet lines independently, which also yields
a robust velocity uncertainty from the r.m.s. scatter of the three resulting velocities. All targets discussed below
have $S/N>2.0$ and $v_{err}<15\kms$. We also measure the $\FeH$ metallicities from the equivalent widths (EW) of the
Ca~II lines \citep{rutledge97}. The metallicities are placed on the \citet{carretta97} scale using the relation
$\FeH=-2.66+0.42 [\Sigma\textrm{Ca}+0.64 (V-V_{HB})]$ where $\Sigma\textrm{Ca}=0.5\textrm{EW}_{\lambda
8498}+1.0\textrm{EW}_{\lambda 8542}+0.6\textrm{EW}_{\lambda 8662}$ and $(V-V_{HB})$ is a surface gravity correction
relative to the $V$-band magnitude of the star, $V$, and the V-band magnitude of the satellite it is assumed to belong
to, $V_{HB}$\footnote{The metallicity of each star in the sample is determined assuming the distance of the satellite it
might belong to. Thus, the derived values are not meaningful if the star is not a satellite member.}. $V$-band
fluxes are calculated from the SDSS colours using transformations derived by our group \citep{ibata07}. $V_{HB}$ is
determined from the colour-magnitude diagrams for the satellite that show a horizontal-branch (Boo, CVnI and UMaI, see
Figure~2) and is fixed as $V_{HB}=m-M+0.7$ for the two others (UMaII and Wil1). Although this is an approximate estimate,
$\FeH$ is not very sensitive to this quantity since $\FeH \propto 0.27(V-V_{HB})$. Previous experience has shown that
$S/N>15$ spectra have uncertainties of $\sim0.2$\,dex on their $\FeH$ value with this method, although this does not
account for systematic shifts that might come from the assumed metallicity scale.

The spectra are then shifted to the rest-frame, and we measure the equivalent width of the Na\textsc{i} doublet lines at
$8183.25\Aa$, and $8194.82\Aa$ by two Gaussian fits (with fixed centroids for the two lines). These lines are gravity
sensitive and can be used to discriminate foreground dwarf stars from the targeted giant stars belonging to the Milky
Way satellites \citep{schiavon97}.

\begin{figure*}
\begin{center}
\includegraphics[angle=270,width=0.49\hsize]{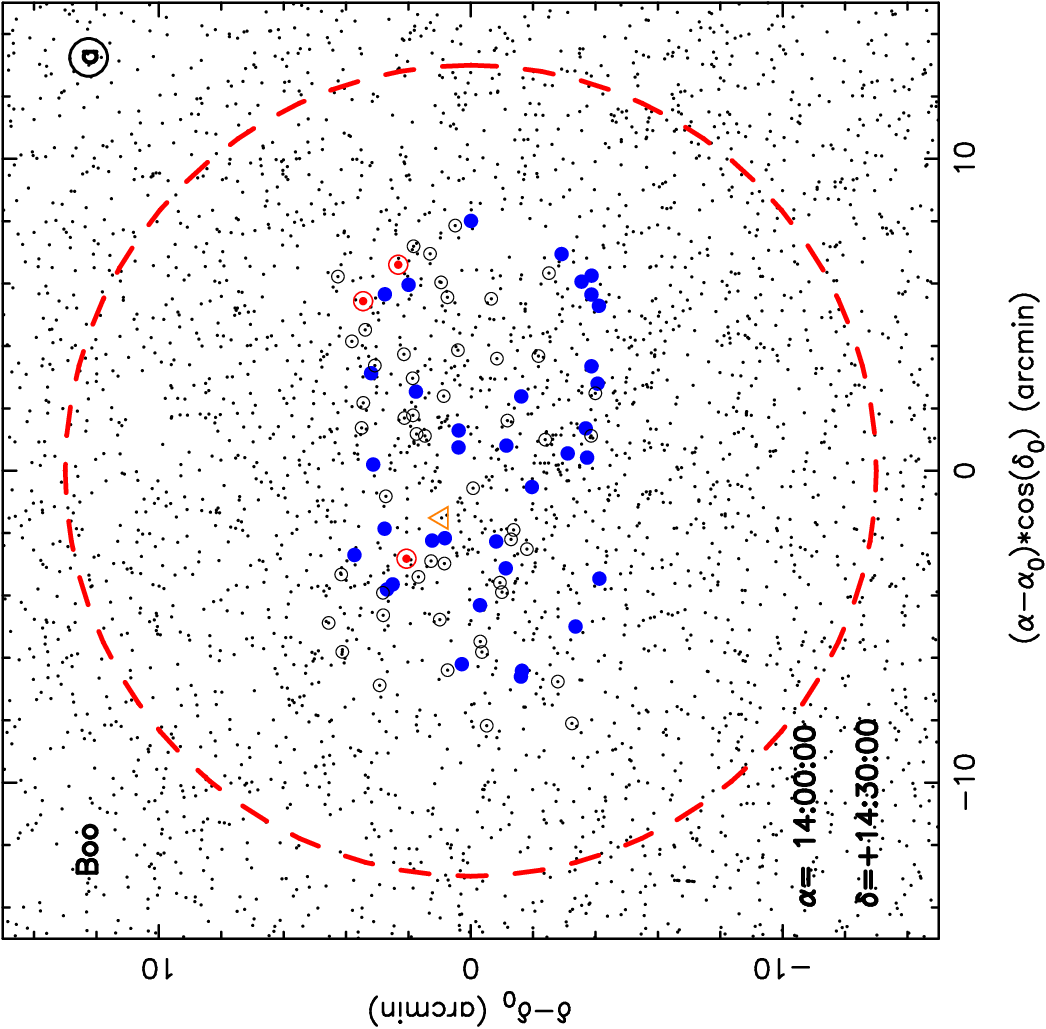}
\includegraphics[angle=270,width=0.49\hsize]{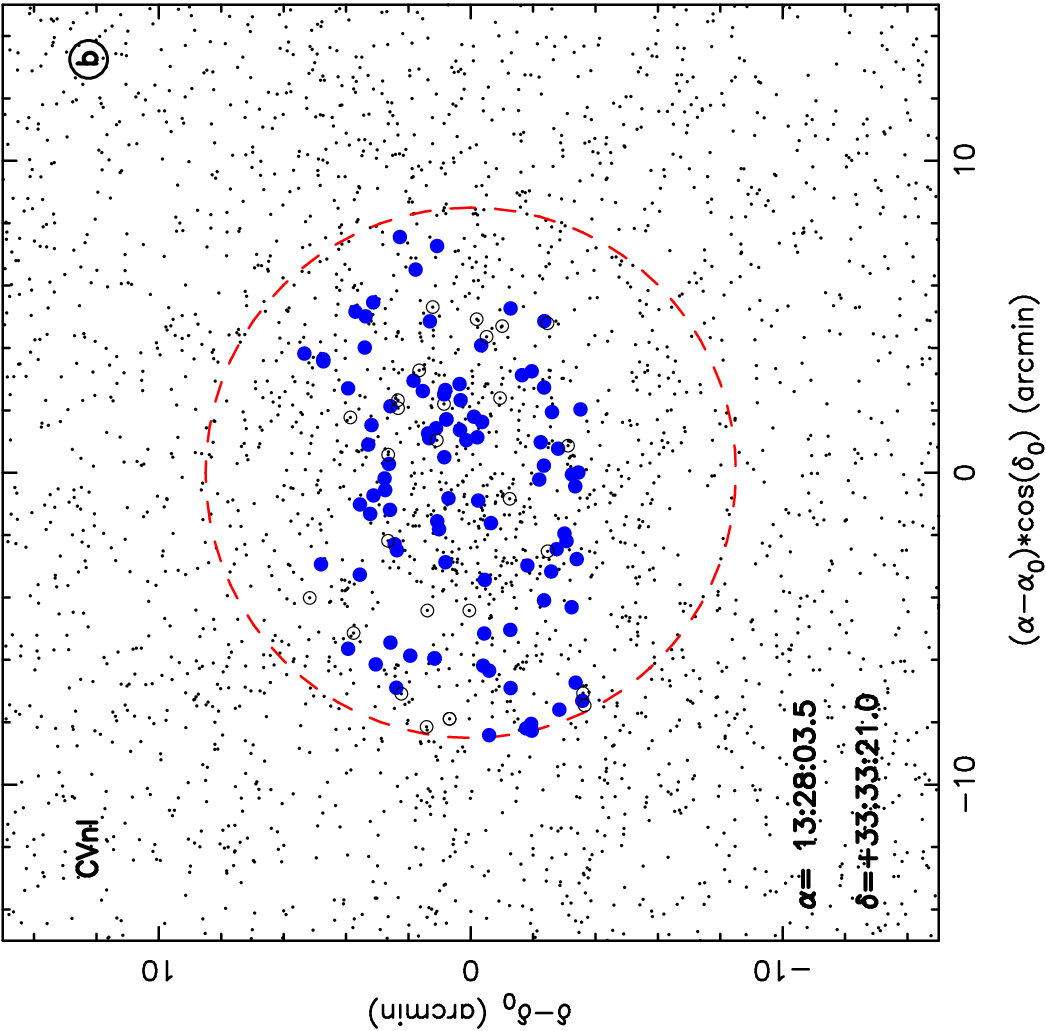}
\includegraphics[angle=270,width=0.49\hsize]{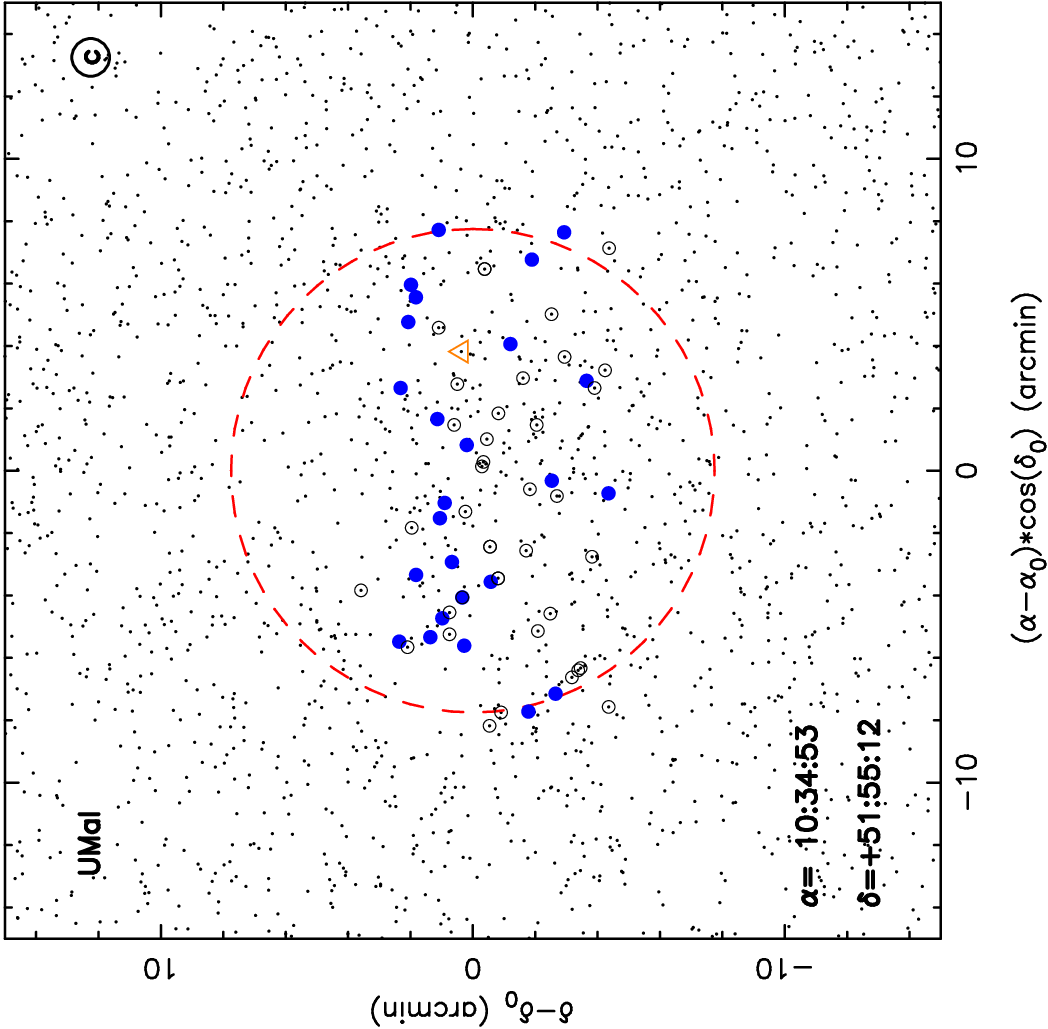}
\includegraphics[angle=270,width=0.49\hsize]{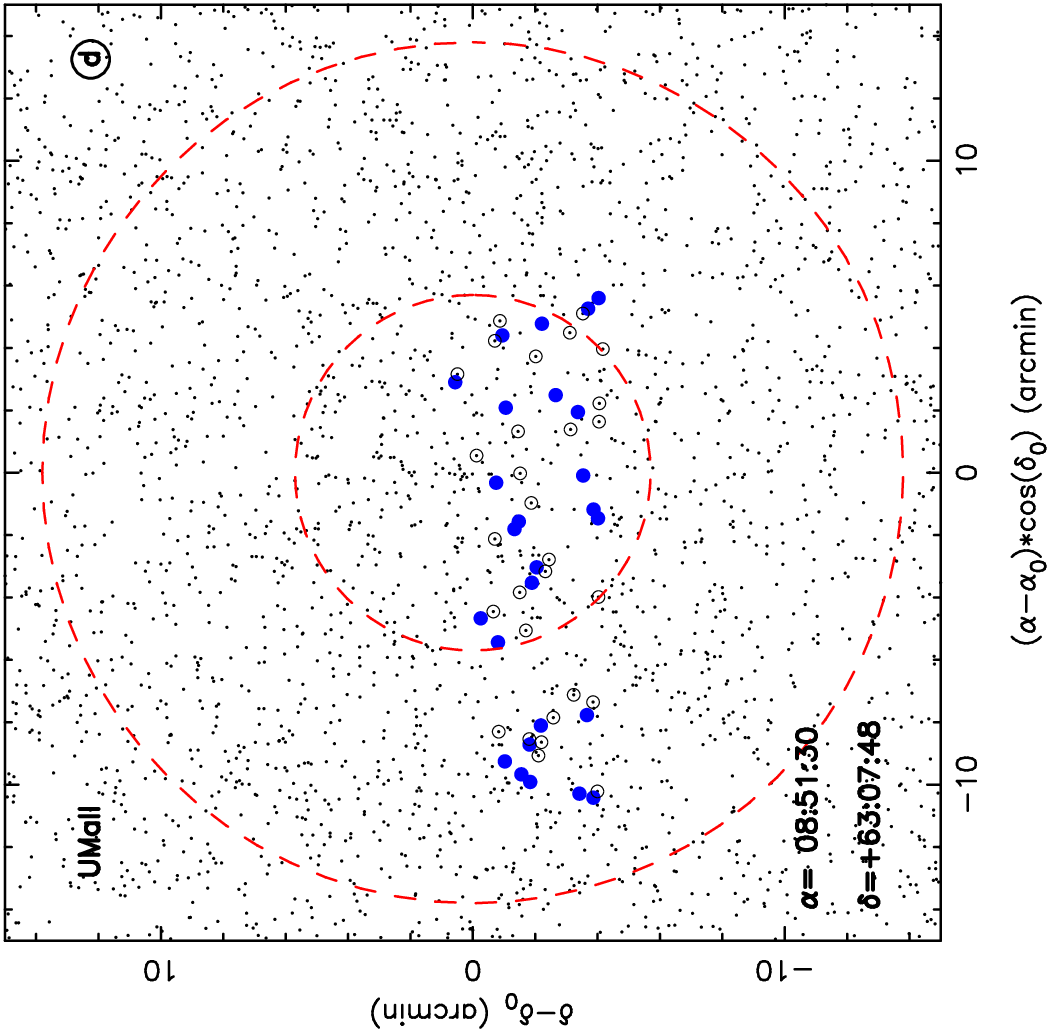}
\caption{The SDSS point-sources around the five Milky Way satellites observed in our survey (\emph{a}: Bo\"otes, \emph{b}: Canes
Venatici~I, \emph{c}: Ursa Major~I, \emph{d}: Ursa Major~II and \emph{e}: Willman~1). Small dots represent all the SDSS
stars in the region, filled circles represent target stars selected from the CMD of the satellites (red giant branch
and main sequence turn-off stars that have a low Na doublet equivalent width; see Figure~2 and~3). Hollow circles
represent field stars that were selected in the same magnitude range or stars that do not pass the Na doublet threshold
and are likely foreground dwarf stars. They are not related to the satellite. In panel \emph{a}, circled big points
represent stars that were initially selected as field stars but have the radial velocity and the CMD position of
Bo\"otes stars; they are included in the analysis to derive Boo parameters. In panels \emph{a} and \emph{c}, triangles
represent stars selected along the RGB of the satellites but that have $\FeH>-1.0$ and are probably not related to the
metal-poor satellites. The dashed lines in all panels correspond to the half-light radius (in the case of UMaII, panel
\emph{d}, the two limits of the half-radius measured by \citealt{zucker06c} are shown). \textit{[This Figure is
available in colour in the online version of the journal.]}}
\end{center}
\end{figure*}

\begin{figure*}
\begin{center}
\includegraphics[angle=270,width=0.49\hsize]{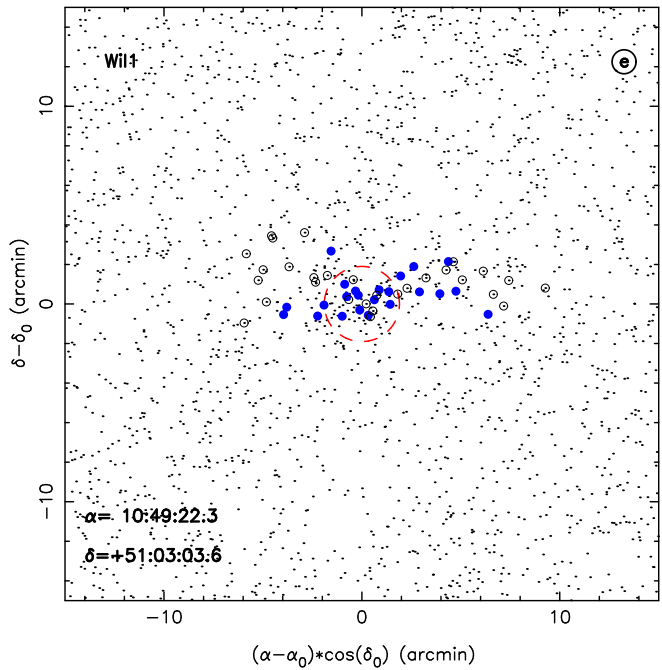}
\addtocounter{figure}{-1}
\caption{\emph{continued}}
\end{center}
\end{figure*}

\begin{figure*}
\begin{center}
\includegraphics[angle=270,width=0.33\hsize]{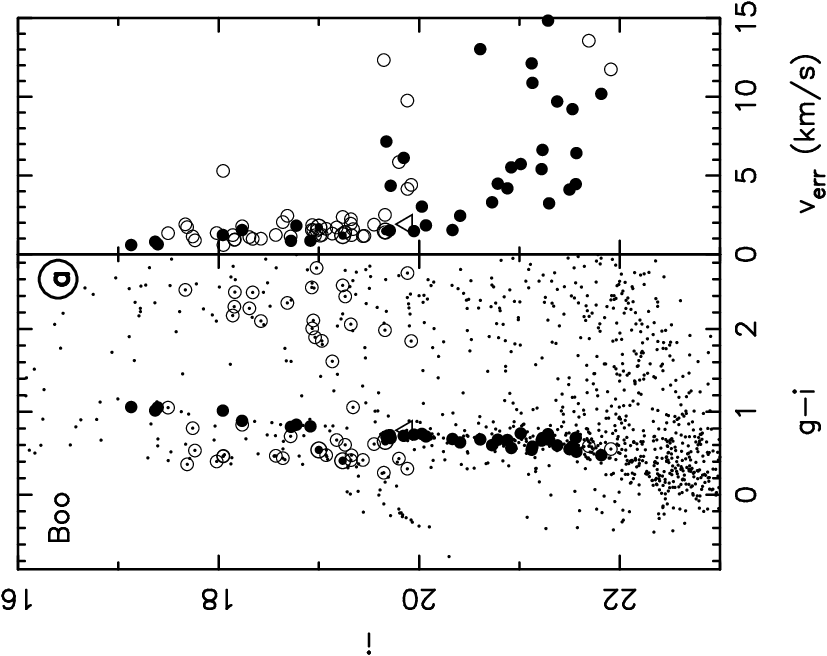}
\includegraphics[angle=270,width=0.33\hsize]{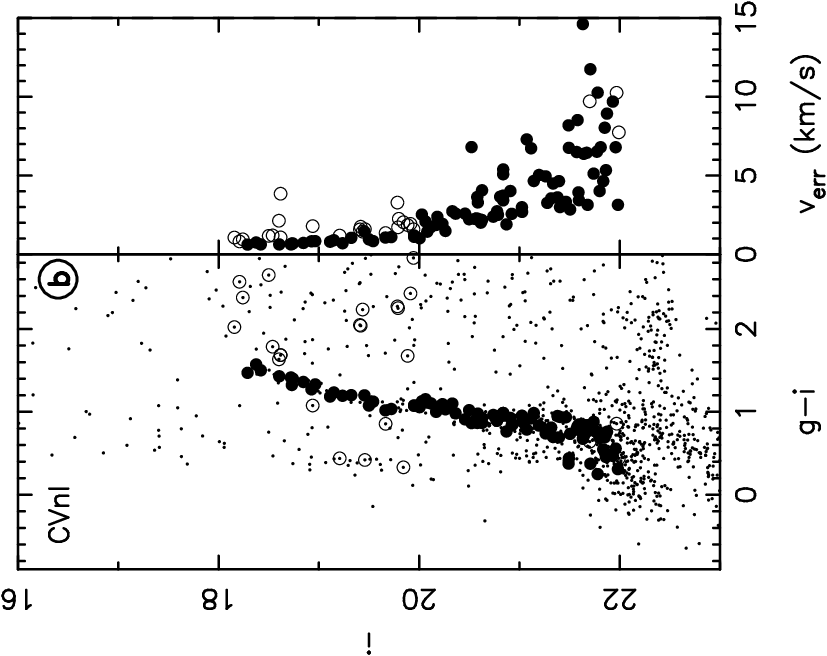}
\includegraphics[angle=270,width=0.33\hsize]{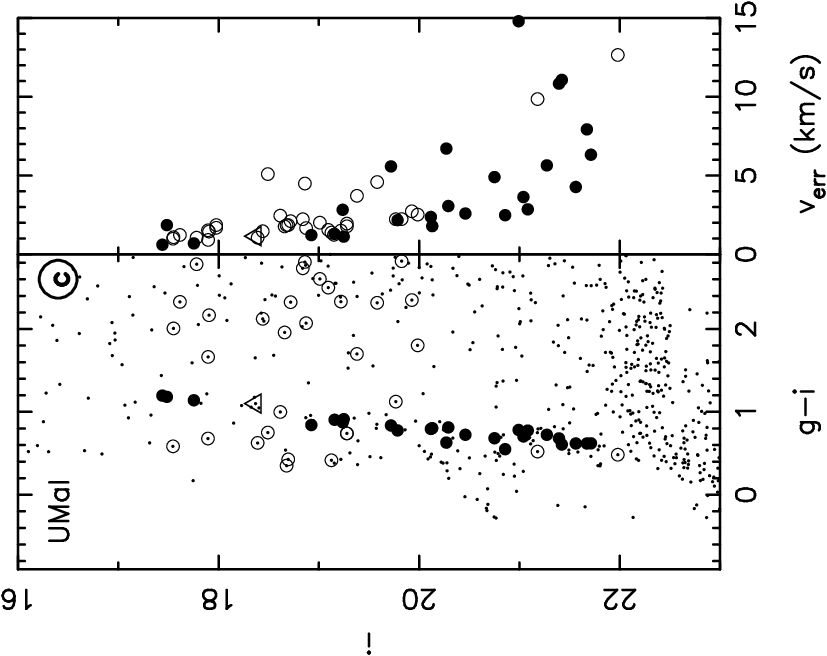}
\includegraphics[angle=270,width=0.33\hsize]{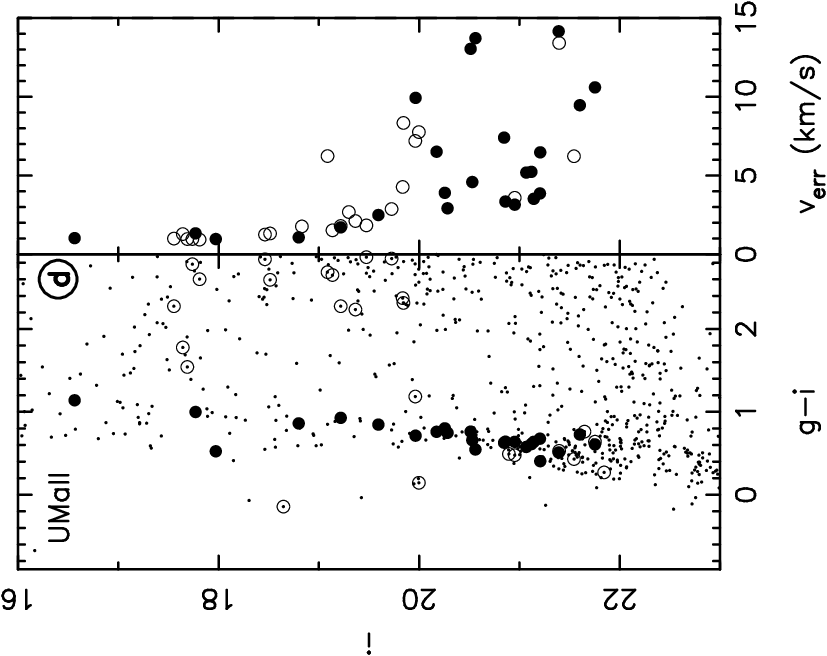}
\includegraphics[angle=270,width=0.33\hsize]{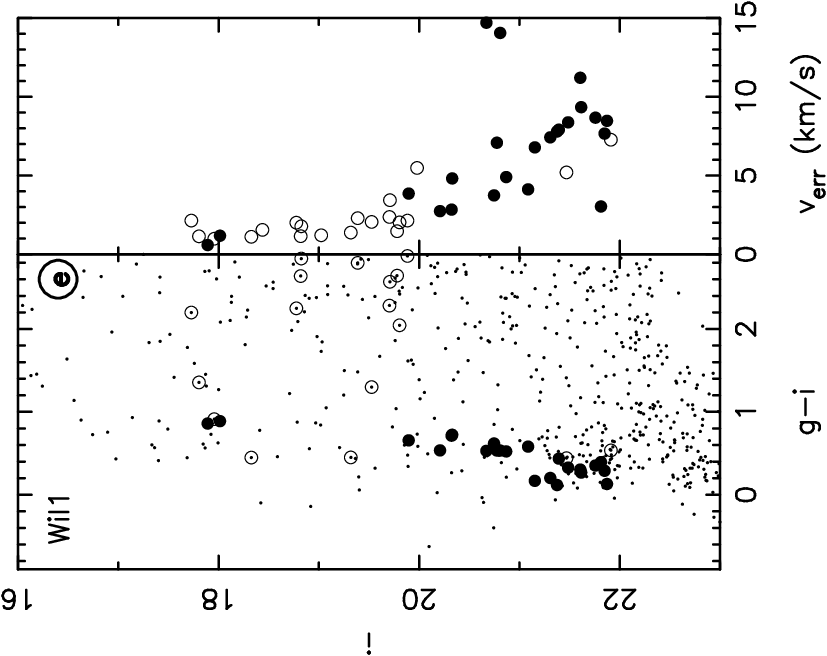}
\caption{SDSS colour-magnitude diagrams within $10'$ of the center of the satellites (left panel of each Figure) and
radial velocity uncertainties of the observed stars in the five satellites in this survey (right panel of each Figure).
As in Figure~1, Figure~2\emph{a} corresponds to Boo, Figure~2\emph{b} to CVnI, Figure~2\emph{c} to UMaI,
Figure~2\emph{d} to UMaII and Figure~2\emph{e} to Wil1. The symbols are the same as in Figure~1.}
\end{center}
\end{figure*}

\begin{figure*}
\begin{center}
\includegraphics[angle=270,width=0.195\hsize]{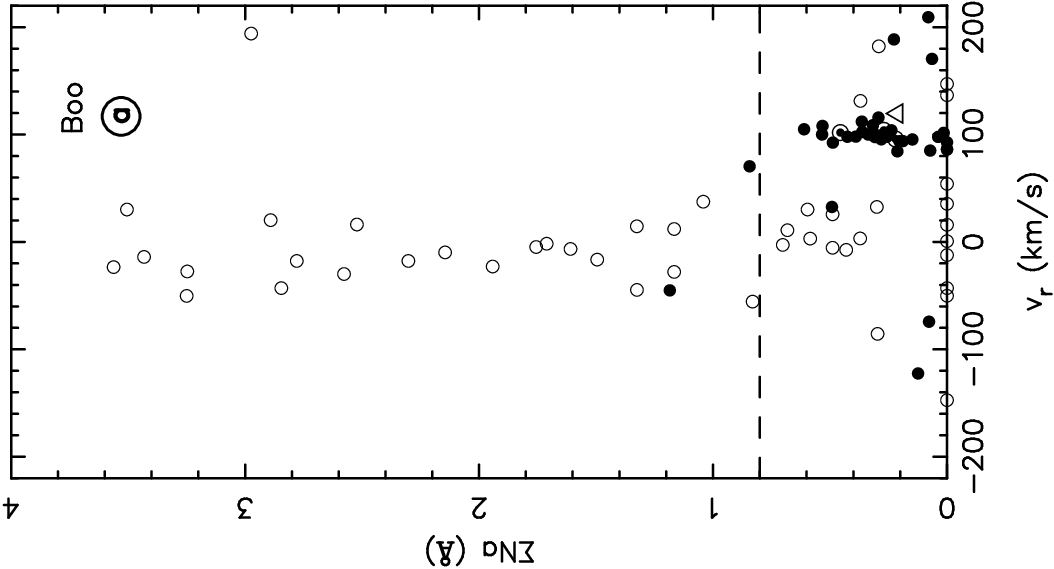}
\includegraphics[angle=270,width=0.195\hsize]{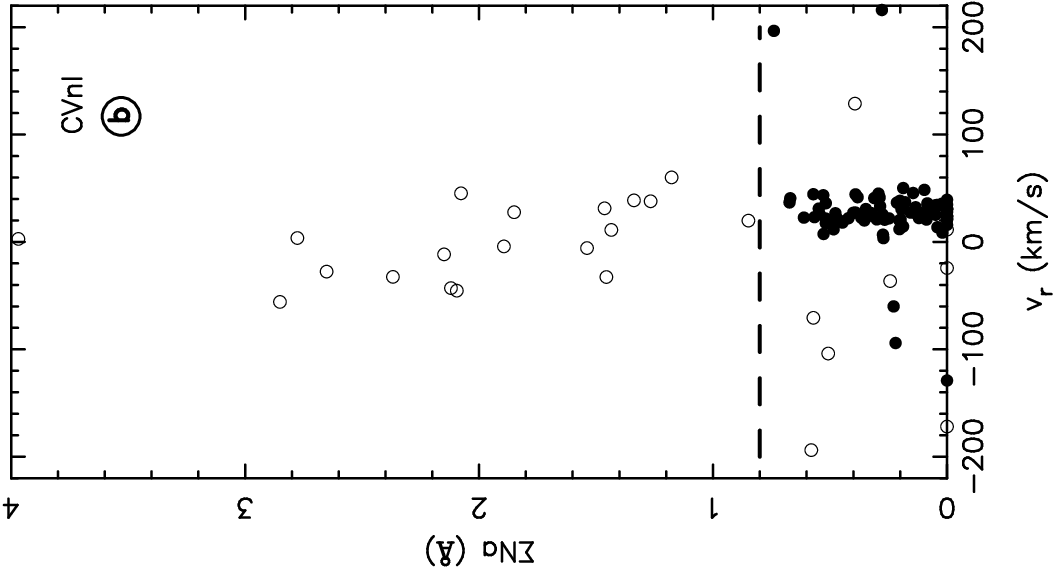}
\includegraphics[angle=270,width=0.195\hsize]{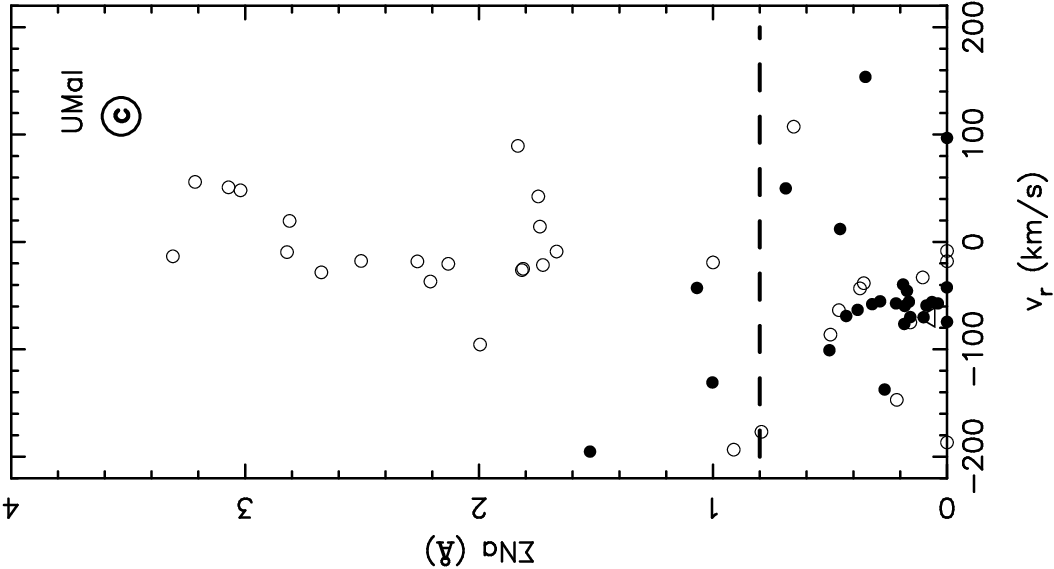}
\includegraphics[angle=270,width=0.195\hsize]{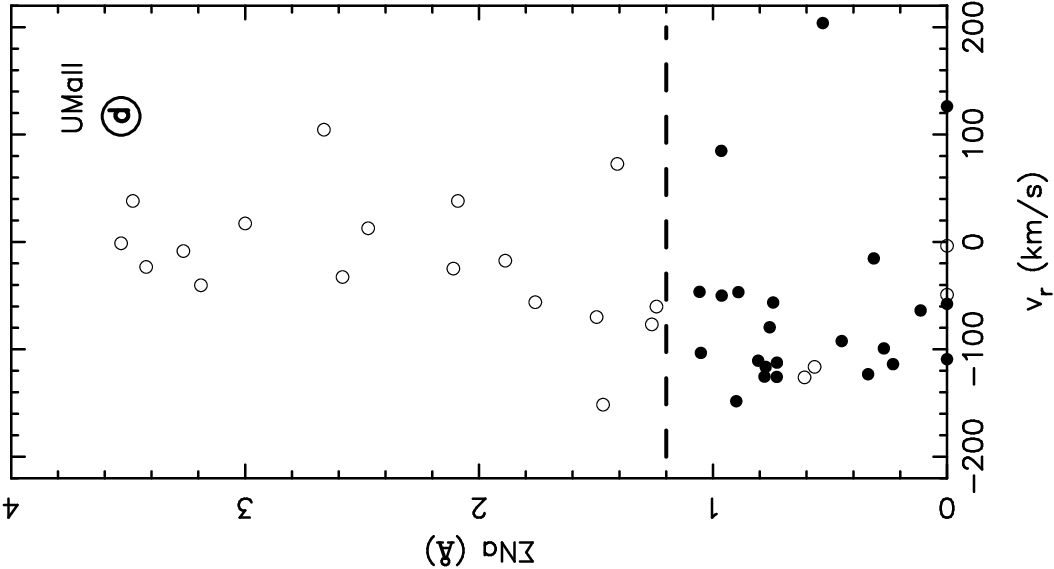}
\includegraphics[angle=270,width=0.195\hsize]{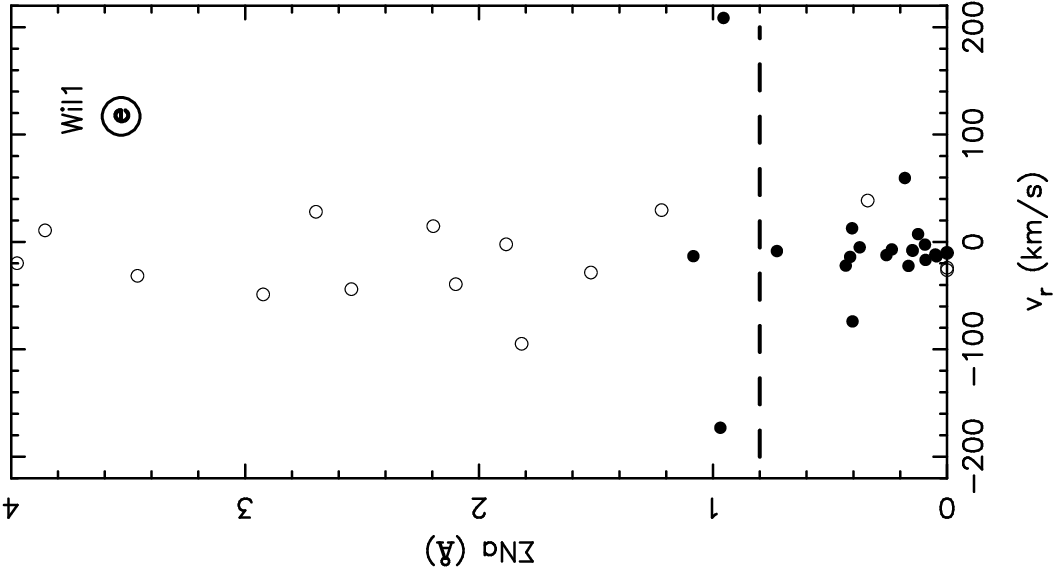}
\caption{Distribution of the sodium doublet equivalent width for the stars observed in each satellite. The symbols are
the same as in Figure~1 except for the filled circles that here represent the initial CMD selection of satellite
members, prior to the $\Sigma\textrm{Na}$ cut. Except in the case of UMaII, the difference between field stars and
satellite members is directly visible and, as expected, giant stars (filled circles) have low $\Sigma\textrm{Na}$
contrary to contaminating dwarfs. From these plots we derive the $\Sigma\textrm{Na}=0.8\Aa$ cut we use to isolate
genuine giants (in the case of UMaII, see section~6 for more details).}
\end{center}
\end{figure*}

In Figure~3, the sum of the Na\textsc{i} equivalent widths,
$\Sigma\textrm{Na}=\textrm{Na}_{\lambda8193}+\textrm{Na}_{\lambda8195}$, is compared for stars selected to belong to the
CMD features of the satellites (red giant branch and horizontal branch) and represented as filled circles and field
stars chosen in the same magnitude range (hollow circles). It is readily visible that satellite stars have much lower
$\Sigma\textrm{Na}$ values than field stars, as is expected for giant stars with lower surface gravity. This effect
is most visible for the populous samples of Bo\"otes and Canes Venatici~I. From these, we will use
$\Sigma\textrm{Na}=0.8$ as the threshold between giant stars belonging to the Milky Way satellites and foreground
contaminating dwarf stars. It remains a perfectly valid threshold for Ursa Major~I and Willman~1 though in both cases
there is one star that was selected along the RGB of the satellite and with a slightly higher $\Sigma\textrm{Na}$ that
is hence not considered as a satellite star. Since in these two cases, the satellite and Galactic contamination radial
velocities overlap, we prefer removing a potential satellite star in order to have a more secure sample. In the case of
Ursa Major~II $\Sigma\textrm{Na}$ does not seem to be as efficient in separating dwarf from giant stars although the
data are of similar quality (see \S~6 for more details).

\section{Bo\"otes}
Bo\"otes was discovered in the SDSS as an overdensity of stars that are aligned in the colour-magnitude
diagram and follow a well-defined red-giant and horizontal branch \citep{belokurov06a}. Although very faint
($M_V=-5.8\pm0.5$, but this value may be underestimated, see \citealt{munoz06}), this new Milky Way satellite has
characteristics typical of dwarf galaxies with a half-light radius of $227\pm13\pc$ \citep{belokurov06a}, a heliocentric
distance of $62\pm3\kpc$ (determined from RR Lyrae stars; \citealt{siegel06}) and is dominated by a metal-poor
population with estimates ranging from $\FeH\sim-2.5$ \citep{munoz06,siegel06} to $\FeH=-2.0$ \citep{siegel06}. A first
spectroscopic analysis of the dwarf galaxy has been performed by \citet{munoz06} and among their 58 targets, the 7~Boo
members that are within the half-light radius yield a systemic velocity and velocity dispersion of
$v_{r,\mathrm{Boo}}=95.6\pm3.4\kms$ and $\sigma_{vr,\mathrm{Boo}}=6.6\pm2.3\kms$ respectively. As is often done for such
studies, they use this velocity dispersion to derive a total mass of $\sim1\times10^7\msun$ for the system, implying a
mass to light ratio that is higher than 100 in solar units.

\begin{figure*}
\begin{center}
\includegraphics[angle=270,width=0.8\hsize]{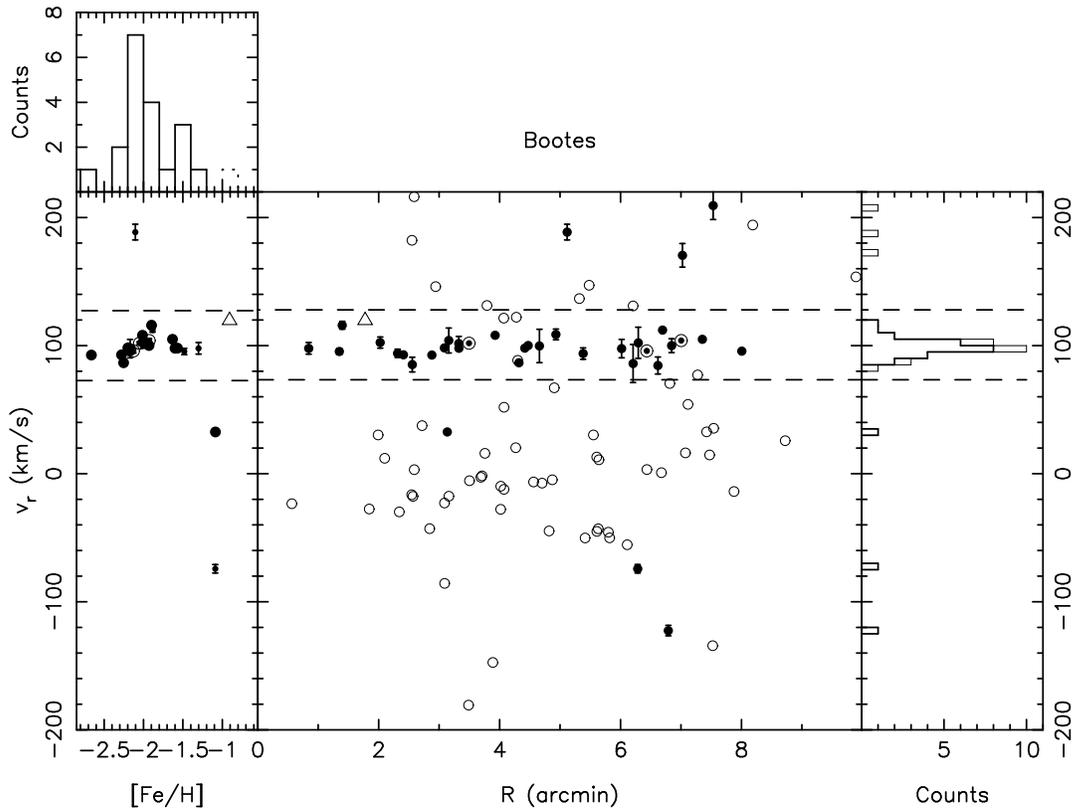}
\caption{Summary of the spectroscopic DEIMOS observations of Bo\"otes. The central panel shows the radial velocity of
each target star as a function of distance to the center of the dwarf galaxy. Hollow circles are stars that were
excluded by the gravity sensitive $\Sigma\textrm{Na}$ cuts and are most likely foreground dwarf stars. Filled circles
represent the most-probable Bo\"otes members, that is, stars aligned along the dwarf red giant branche that
were not excluded by the $\Sigma\textrm{Na}$ cut. The triangle also corresponds to an object within the CMD selection
region but that is probably not a Boo member since it has a high metallicity ($\FeH>-1.0$). The three field stars
with Boo-like velocities are represented by circled big dots and are considered as Boo members since they are also
located close to the CMD features of the dwarfs. The error bars represent the $1\sigma$ radial velocity uncertainties,
those that are not shown are smaller that the symbols and lower than $\sim5\kms$. The thin line in the right panel
represents the radial velocity distribution of the Boo-members whereas the thick line is restricted to those stars with
an uncertainty lower than $6\kms$. The bottom left panel presents the metallicity distribution for these stars that also
have $S/N>15$ as big filled circles necessary to derive a reliable $\FeH$ value (with uncertainties of $\pm0.2$~dex) and
these with $15>S/N>10$ and less reliable $\FeH$ as smaller dots. There is no significant difference between the two
sub-samples. In the three bottom panels, the dashed lines correspond to the radial velocity limits that were used to
isolate Boo star members (see the text for more details). The top left panel represents the metallicity distribution of
the Bo\"otes members, that is, stars that form the radial velocity peak. The dotted histogram includes the star that has
$\FeH>-1.0$ and that is probably not a member of the dwarf galaxy.}
\end{center}
\end{figure*}

Figure~4 summarizes the spectroscopic information of our Bo\"otes sample observed with DEIMOS and Table~1 lists the
individual data on each observed star. In the central panel of the Figure, there is a clear separation between
foreground dwarf stars with $\Sigma\textrm{Na}>0.8$ (hollow circles) and stars that are aligned along the CMD features
of the dwarf galaxy (filled circles) which are clustered around $\sim100\kms$ as expected. However, three of the field
stars fall within the Boo peak. Since they are also close to the RGB and HB of the dwarf galaxy and have
$\Sigma\textrm{Na}<0.8$, we consider them as Boo members (the circled big dots of Figure~2a, 3a and 4). One of the stars
in the sample, represented by a triangle in Figure~4, has a radial velocity that is close to the Boo peak but is more
metal-rich than the other Boo stars. A signal-over-noise ratio of 21 makes it unlikely that its metallicity has
high uncertainties. Other stars in the sample with $\FeH\sim-1.0$ are clearly not Boo members, which also hints at a
$\FeH$ value that is not meaningful for this star. However, should later studies reveal that Boo contains
such a metal-rich population, we will determine the velocity parameters of the dwarf with and without this star.

Using only the confirmed Boo stars, we iteratively determine the mean radial velocity, $v_r$, and velocity dispersion,
$\sigma$, of the dwarf galaxy by clipping stars within $\pm3\sigma$ of $v_r$ and determine anew the mean radial
velocity and the velocity dispersion of the sample defined in this way. After each iteration, the best parameters are
determined with a maximum-likelihood algorithm that explores a coarse grid of the $(v_r,\sigma)$ space and searches for
the couple of parameters that maximizes the $ML$ function defined as:
\begin{equation}
ML(v_r,\sigma)=\sum_{i=1}^{N}\frac{1}{\sigma_{\mathrm{tot}}}
\exp\Big[-\frac{1}{2}\big(\frac{v_r-v_{r,i}}{\sigma_{\mathrm{tot}}}\big)^2\Big]
\end{equation}
\noindent with $N$ the number of stars in the sample, $\sigma_{\mathrm{tot}}=\sqrt{\sigma^2+v_{err,i}^2}$, $v_{r,i}$ the
radial velocity measured for the $i^\mathrm{th}$ star and $v_{err,i}$ the corresponding uncertainty. Using this
definition of $\sigma_{\mathrm{tot}}$ allows us to disentangle the intrinsic velocity dispersion of the dwarf galaxy
population, $\sigma$, and the contribution of the measurement uncertainties to the observed distribution. By definition,
this technique also gives a low weight to stars with poorly determined radial velocity and is therefore applicable to
the whole velocity sample without removing those stars that have a high velocity uncertainty.

Starting values are taken from \citet[][$v_r=95.6\kms$ and $\sigma=6.6\kms$]{munoz06} and convergence is achieved for
$v_r=99.0\pm2.1\kms$ (or $v_{\mathrm{gsr}}=106.5\pm2.1\kms$) and $\sigma=6.5_{-1.4}^{+2.0}\kms$ determined with a final
sample of 30~stars. This is in good agreement with but more precise than the value of \citet{munoz06} determined from
only 7~stars. Restricting the sample to the 24~stars with $v_{err}<6\kms$ (the thick line in the right panel of
Figure~4)\footnote{The $6\kms$ threshold is a good compromise between the quality of the individual velocity values and
the number of stars in the sample.} to ensure that the individual uncertainties do not influence the convergence, the
derived parameters are statistically equivalent ($\sigma=6.5_{-1.3}^{+2.1}\kms$ and $v_r=99.9\pm2.4\kms$). With this
quality cut, the radial velocity distribution that is represented by the thick line in the right panel of Figure~4 shows
a very clear peak. In all cases, the stars that are removed by the $3\sigma$ clipping all have radial velocity that are
clearly different from the systemic velocity of Boo. Adding the metal-rich star yields a slightly higher velocity dispersion
$\sigma=7.4_{-1.2}^{+2.2}\kms$ centered on $v_r=99.8\pm2.4\kms$.

As is usually done for dwarf galaxies, one can try to estimate the mass of Boo from its central velocity dispersion, and
structural parameters, assuming it is a spherical system in virial equilibrium. Even though it may not be the case, the
central velocity dispersion can still be used to measure the instantaneous mass content of the structure
\citep[e.g.][]{oh95,piatek95}. The \citet{richstone86} formula can be used to derive the mass-to-light ratio $(M/L)$ of
the system in solar units:
\begin{equation}
M/L=\eta\frac{9}{2\pi G}\frac{\sigma_0^2}{S_0 r_{hb}} 
\end{equation}
\noindent where $\eta$ is a scale parameter close to 1.0, $S_0$ is the central surface brightness of the dwarf, $r_{hb}$
its half-light radius and $\sigma_0$ its central velocity dispersion. Alternatively, the \citet{illingworth76} approach
directly gives the mass $M$ of the system in solar units:
\begin{equation}
M=167\,r_c\,\mu\,\sigma_0^2
\end{equation}
\noindent with $r_c$ the core radius of the system in parsecs and $\mu$ a scale factor taken as 8 by \citet{mateo98}.
Both methods are not highly accurate since for equation (2), $S_0$ can have uncertainties of $\sim50\%$
for the faint systems considered here and determining the mass requires using the luminosity of the system, itself not
well constrained (see e.g. \citealt{munoz06}). Equation (3) should yield a better mass estimate since it does not
require the (poorly constrained) luminosity of the satellites as an input parameter but it relies on the core radius
$r_c$ of the systems, which have not been determined with accuracy for the satellites studied in this paper. Following
\citet{munoz06}, we use the approximation $r_c\sim r_{hb}$ which is usually reliable within $\pm25\%$ although there are
some dwarf galaxies which show sizable difference between $r_c$ and $r_{hb}$ (see e.g. \citealt{mcconnachie06}).

Applying Equation (2) for Bo\"otes with $\sigma_0=\sigma=6.5\kms$, $S_0\sim0.20\lsun\pc^{-2}$ derived
from the exponential profile of the dwarf determined by \citet{belokurov06a} yields $(M/L)=220$ in solar units.
Assuming $L=1.8\times10^4\lsun$ ($M_V=-5.8$) then yields a total mass of $M\sim4\times10^6\msun$ although if we follow
\citet{munoz06} and use the brighter $L=8.6\times10^4\lsun$ ($M_V=-6.75$) we obtain $M\sim1.8\times10^7\msun$. On the
other hand, Equation (3) yields $M\sim1.3\times10^7\msun$, obviously in agreement with \citet{munoz06} since the
velocity dispersion we use is very similar to theirs.

Bo\"otes is close enough to the Milky Way that a significant portion of the observed stars have a high enough S/N to
allow for a proper determination of their metallicity. The metallicity distribution of the most probable Boo stars is
presented on the bottom left panel of Figure~4 with the values derived from the highest quality spectra ($S/N>15$ or
0.2\,dex uncertainty on $\FeH$) as the bigger filled circles and lower quality spectra ($15>S/N>10$) as smaller filled
circles. There is no significant difference in the distribution of these two sub-samples and we therefore merge them
together to study the metallicity of Boo. Isolating stars in the velocity peak yields the distribution of the top left
panel and confirms Boo is a metal-poor satellite of the Milky Way, though somewhat less metal-poor than was measured in
previous studies based on isochrone fitting with SDSS photometry \citep[$\FeH\sim-2.3$;][]{belokurov06a}, a combined
spectrum of 7 confirmed Boo stars \citep[$\FeH\sim-2.5$;][]{munoz06} or RR Lyrae
\citep[$\FeH\sim-2.5$;][]{siegel06}. The median metallicity derived from the DEIMOS sample is indeed $\FeH=-2.1$, though
one star is as metal-poor as $\FeH=-2.7$. It is unclear why the value is more metal-rich than previous estimates. A
comparison with metallicities determined photometrically by comparison of the RGB stars with the 10 Gyr isochrones
from \citet{girardi04} shows only a tiny systematic shift to more metal-rich values of  0.1\,dex (in agreement with the
\citealt{belokurov06a} photometric metallicity estimate) but this shift could also be due to the age assumed with the
isochrones. The discrepancy in the metallicity may also be related to the \citet{carretta97} metallicity scale used here
compared to the \citet{zinn84} scale used by \citet{siegel06}. Figure~5 of \citet{carretta97} points out that the Zinn \&
West scale is always $\sim0.1$\,dex more metal-poor than the Carretta \& Gratton scale for $\FeH\lta-2.0$ which bridges
part of the discrepancy  with the \citet{siegel06} values. Finally, since none of these systematics are large enough to
account for the 0.5\,dex difference between our value and those measured by \citet{munoz06} and \citet{siegel06}, it has
to be noted that none of the techniques used to estimate $\FeH$ have been calibrated for $\FeH\lta-2.3$. The
extrapolated relations used in all these studies (including ours) could therefore start diverging at the low metallicity
of Boo.

\section{Canes Venatici~I}
Discovered in the SDSS by \citet{zucker06b}, Canes Venatici~I is the furthest of the newly
discovered Milky Way faint dwarf galaxies with a heliocentric distance of $\sim220\kpc$. It is also the brightest, with
$M_V=-7.9\pm0.5$, making it a galaxy similar to the Draco dSph, but with a bigger physical size (half-light radius
$r_h\sim550\pc$). The populated red giant branch of this satellite (see Figure~2\emph{b}) makes it a perfect target for
DEIMOS and we were able to observe 127~stars with two pointings. An analysis of this dwarf galaxy is presented in
details in \citet{ibata06} and reveals that CVnI hosts two distinct stellar populations: one is metal-poor
($-2.5\lta\FeH\lta -2.0$), kinematically hot ($\sigma=13.9_{-2.5}^{+3.2}\kms$) and extended
($r_{hb}=7\mcnd8_{-2.1}^{+2.4}$) while the second one is concentrated at the center of the dwarf
($r_{hb}=3\mcnd6_{-0.8}^{+1.1}$), is more metal-rich ($-2.0\lta\FeH\lta -1.5$) and has a near-zero velocity dispersion
(with a dispersion lower than $1.9\kms$ at the 99\% confidence level). Though there is growing evidence that many dwarf
galaxies harbor kinematically and structurally distinct populations or at least population gradients (see e.g.
\citealt{harbeck01}; \citealt{tolstoy04} for Sculptor; \citealt{battaglia06} for Fornax), CVnI is to date the most
extreme case of such a behaviour. It is also the first time that a near-zero velocity dispersion is discovered in a
dwarf galaxy though deeper observations are required before putting strong constraints on the dark matter content of this
dwarf galaxy.

We nevertheless reproduce Figure~4 for the Canes Venatici~I dwarf in Figure~5 for comparison and list the
characteristics of the observed stars in Table~2

\begin{figure*}
\begin{center}
\includegraphics[angle=270,width=0.8\hsize]{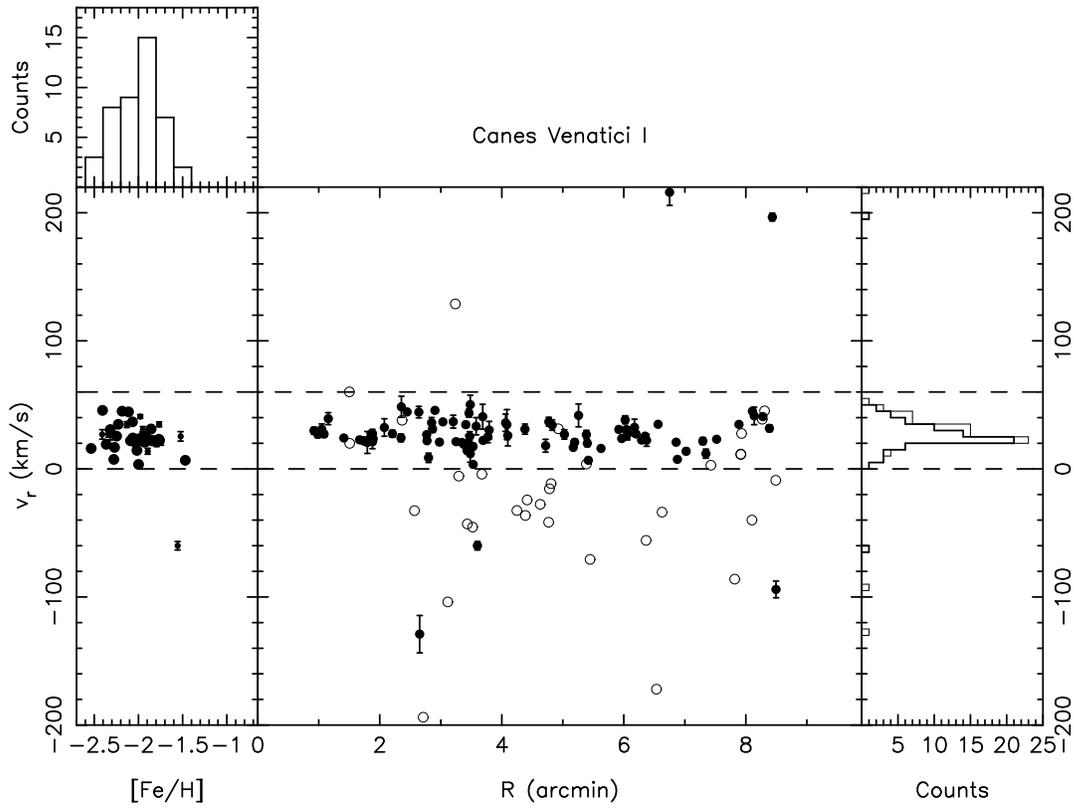}
\caption{Same as Figure~4 for the Canes Venatici~I dwarf galaxy.}
\end{center}
\end{figure*}

\section{Ursa Major I}
Ursa Major I was discovered by \citet{willman05b} from an automatic search of faint substructures
in the SDSS. Though the parameters of this satellite are not well constrained given the low number of stars on its red
giant branch and the absence of deep observations that reach the more populated main sequence turn-off, UMaI appears to
be at a distance of $\sim100\kpc$, with $M_V\sim-6.75$, $r_{hb}=250\pc$ and a metallicity roughly estimated to be
within $-2.1\lta\FeH\lta-1.7$. \citet{kleyna05} observed 7~stars near the tip of the red giant branch with the HIRES
spectrograph mounted on the Keck~I telescope and concluded that 5 of these are likely members of the galaxy. They derive
a mean systemic velocity of $v_r=-52.4\pm4.3\kms$ and a velocity dispersion of $\sigma=9.3_{-1.2}^{+11.7}\kms$ through a
maximum-likelihood algorithm and subsequently use these values to infer a mass-to-light ratio of the order of
$\sim500$ in solar units for UMaI.

To increase the number of observed stars in the dwarf galaxy, two DEIMOS fields were observed at the center of UMaI,
roughly covering the region within the half-light radius of the dwarf (Figure~1\emph{c}). As before, Figure~2\emph{c}
shows the CMD taken from the SDSS for this region of the sky along with the selected targets and the Na\textsc{i}
doublet discrimination is presented Figure~3\emph{c}. The parameters of the observed stars are listed in Table~3. Two of our
stars were observed by \citet{kleyna05} with the HIRES spectrograph on Keck (their stars 1 and 7). The shift between the
two measurements of star~1 is small ($2.3\kms$) whereas star~7 shows a shift of $13.1\kms$, but this is probably due to
the high uncertainty of the HIRES value ($>5\kms$) since the DEIMOS observation has an uncertainty of only $1.2\kms$.
As there are only three stars in their sample that were not re-observed with DEIMOS, we prefer not to add them to our $\sim17$
stars that fall in the velocity range of UMaI to avoid adding a source of systematics.

\begin{figure*}
\begin{center}
\includegraphics[angle=270,width=0.8\hsize]{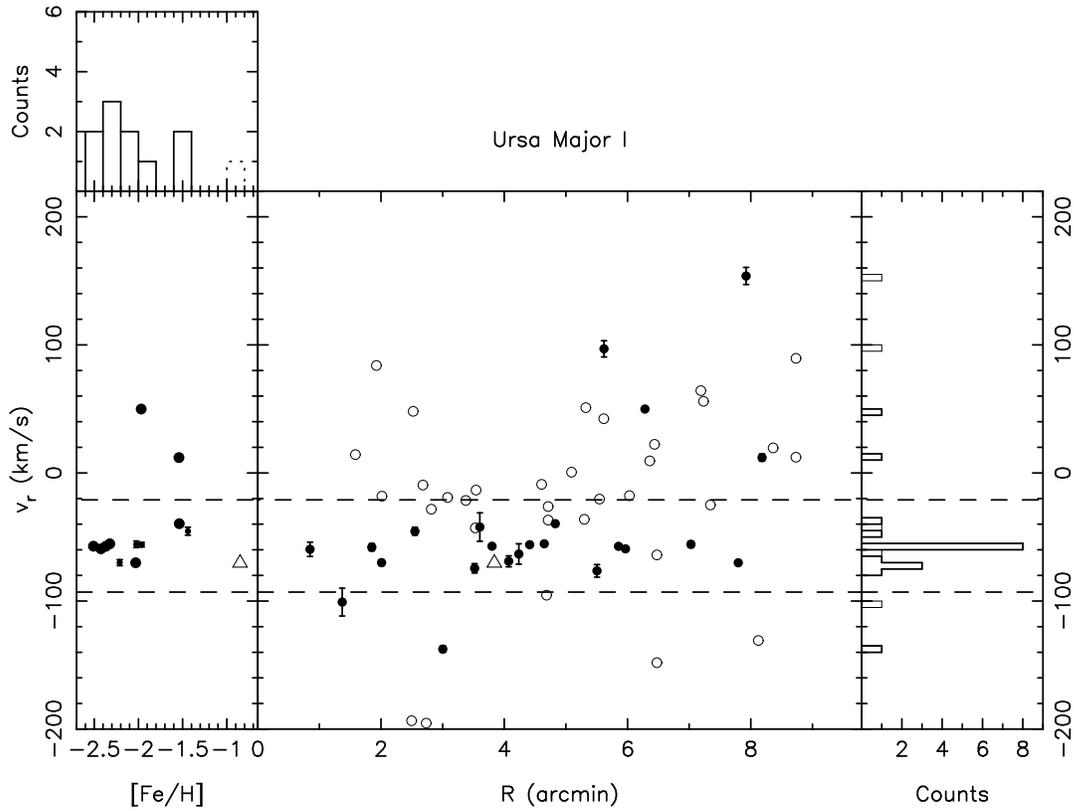}
\caption{Same as Figure~4 for the Ursa Major~I dwarf galaxy.}
\end{center}
\end{figure*}

Contrary to Bo\"otes, Ursa Major I has a velocity distribution that overlaps with the Galactic contaminants (Figure~6).
The dwarf galaxy stars however seem to produce a very narrow peak of 8~stars surrounded by a broader group of stars for
which it is difficult to definitely conclude whether they belong to UMaI or to the Galactic contamination (only one star
is removed from the sample since, with a metallicity of $\FeH=-0.8$, it is inconsistent with the metal-poor population
of the dwarf and is hence most probably a dwarf for which the determined $\FeH$ has no meaningful sense). Applying
the maximum likelihood technique that was applied on the Boo sample yields a systemic velocity of $v_r=-57.0\pm3.5\kms$
and a velocity dispersion $\sigma=11.9_{-2.3}^{+3.5}\kms$ for the stars aligned along the CMD features of the dwarf, in
good agreement with the \citet{kleyna05} measurement\footnote{Including the metal-rich star leads to a systemic
velocity $v_r=-59.1\pm4.1\kms$ and a velocity dispersion $\sigma=12.8_{-2.5}^{+3.6}\kms$.}. A Kolmogorov-Smirnov test
between the 17\,stars used for this best fit and the fit itself however only gives a probability of 19\% that the data
follow the fit distribution. With this in mind, it is interesting to note that the narrow peak of the velocity
distribution at $v_r\sim57\kms$ is reminiscent of the very low velocity dispersion peak observed in CVnI
\citep{ibata06}. This is also of particular interest since some of the more broadly-distributed objects may only be
Galactic contaminants as is hinted at by the higher metallicity of the two stars at higher velocity than the peak
($\FeH\sim-1.5$). As for CVnI, this narrow peak is compatible with a dispersion of $0\kms$. It also has a dispersion
that is lower than $3.4\kms$ at the 99\% confidence level.

One can nevertheless use the derived dispersion values to put some constraints on the mass-to-light ratio of UMaI. Since
no central surface brightness value $S_0$ has been measured for UMaI, \citet{kleyna05} has assumed a uniform luminosity
distribution within the half-light radius of the galaxy; this yields $S_0=0.11\lsun\pc^{-2}$ and is certainly a lower
limit to $S_0$ given that dwarf galaxies have peaked luminosity profiles. Indeed, reproducing this estimate for the
Bo\"otes galaxy yields the same value whereas the central surface brightness of this galaxy is measured to be
$S_0=0.20\lsun\pc^{-2}$ when using the central surface brightness measured by
\citet[][see \S~4]{belokurov06a}. Hence, we prefer using $S_0\sim0.20\lsun\pc^{-2}$ for UMaI and, along with
$\sigma=11.9\kms$ and $r_{hb}=250\pc$, we derive $(M/L)\sim900\msun/\lsun$ using equation (2). With
$L\sim4.3\times10^4\lsun$, this converts to $M\sim3.8\times10^7\msun$. Alternatively,
$M=4.7\times10^7\msun$ can be derived from equation (3). These estimates make UMaI a highly dark
matter dominated dwarf galaxy. However, as was shown by \citet{ibata06} in the analysis of CVnI, deriving the total mass
of the system may not be straightforward if the dwarf galaxy hosts more than one stellar component. If we were to use only
the stars in the narrow peak of UMaI, the total mass of the system would be lower than $3.8\times10^6\msun$ at the 99\%
confidence limit through equation (3), a drastically dissimilar value that urges for a larger sample of high accuracy
radial velocities. This could be obtained by observing stars at slightly fainter magnitudes than those presented here
(see Figure~2\emph{c}).

Finally, the left panels of Figure~6 confirm that UMaI is dominated by metal-poor stars with a median metallicity of
$\FeH\sim-2.0$ although if the two most metal-rich stars that have a clear offset in metallicity and are also shifted to
higher velocities than the UMaI peak are removed from the sample, this value shifts to a more metal-poor
$\FeH\sim-2.4$. As for Boo, we have used all stars with $S/N>10$ to derive the metallicity of UMaI since lower quality
spectra ($15>S/N>10$, smaller filled circles in the left panel) do not produce significantly different $\FeH$ values than
higher quality spectra ($S/N>15$, bigger filled circles).

\section{Ursa Major II}

The very faint Ursa Major II satellite is presented in \citet{zucker06c} and was also discussed by
\citet{grillmair06}. With a total magnitude of only $M_V\sim-3.8$, it would be one of the faintest known Milky Way
satellites, though its structural parameters are not well constrained, with a distance of $30\pm5\kpc$ and a half-light
radius that could be between $50\pc$ and $120\pc$. UMaII is peculiar in the sense that it lies spot-on the extrapolation
of the orbit of the so-called Orphan Stream of \citet{belokurov06b} and \citet{grillmair06}, a stellar stream that
extends over $\sim60\deg$ on the sky at distances ranging from $\sim15\kpc$ at its nearest extension to $30\kpc$ where
it is nearest to UMaII (located $\sim10\deg$ away). UMaII also overlaps with the detection of Complex A, an H\textsc{i}
high velocity cloud. The distance to the cloud, estimated between 4 and $10\kpc$ however does not favor a direct link.

\begin{figure*}
\begin{center}
\includegraphics[angle=270,width=0.8\hsize]{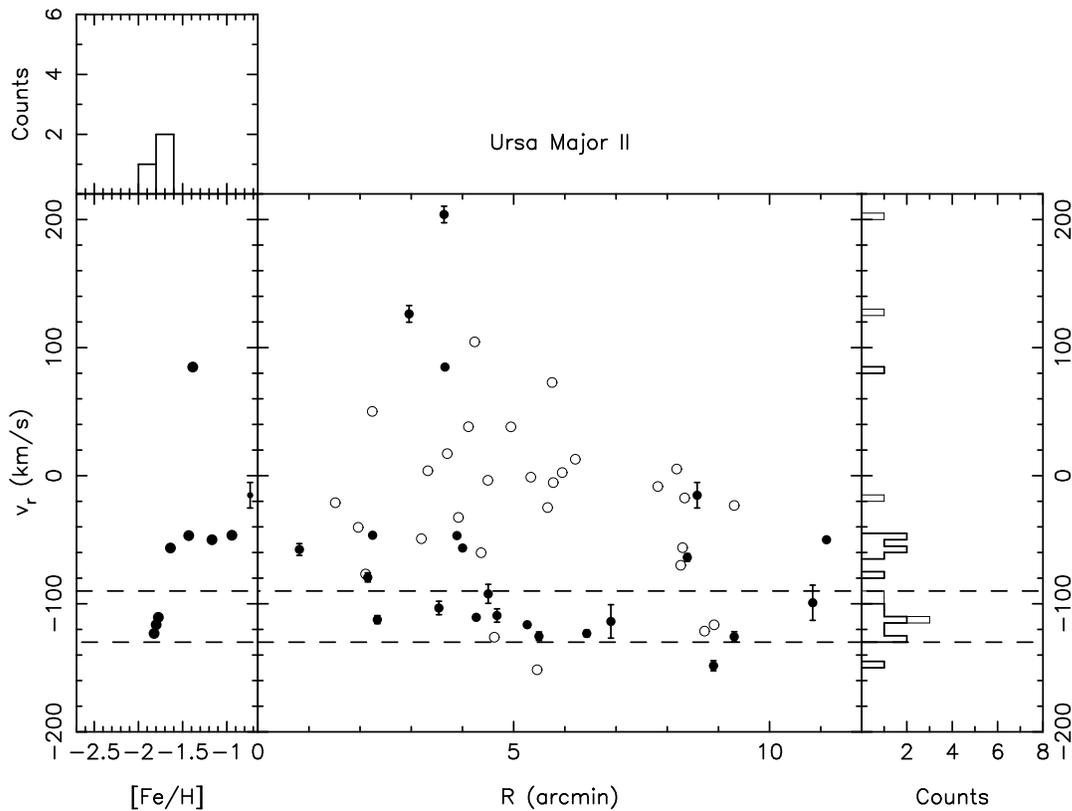}
\caption{Same as Figure~4 for the Ursa Major~II satellite.}
\end{center}
\end{figure*}

One DEIMOS field was targeted at the center of the object (see Figure~1\emph{d} where the two dashed circles
represent the two boundaries of the half-light radius estimate). Although the spectra are not of lower quality than
those observed for the other galaxies, the Na\textsc{i} doublet does not seem as efficient as for the other
samples where the stars selected along the giant branch of the galaxies are clearly clumped at low values of
$\Sigma\textrm{Na}$ (Figure~3). To avoid removing genuine UMaII stars, we relax the cut and only consider stars with
$\Sigma\textrm{Na}>1.2$ as foreground dwarf stars. In this way, we keep all but one stars aligned along the RGB of
UMaII. The radial velocity distribution of these stars is presented on Figure~7 and shows no clear radial velocity peak
(the stars parameters are listed in Table~4). The most probable UMaII members (filled circles) are distributed in two
small groups of stars : one group has $v_r\sim-120\kms$, and is composed of stars with a similar metallicity
($\FeH\sim-1.8$). Since it is clearly distinct from Galactic contaminants (hollow circles), and has the preferred
metallicity value of \citet{zucker06c} for UMaII, it seems more likely that this group of stars belongs to the UMaII
structure. Therefore, we henceforth only consider the group of stars with the lowest radial velocity as UMaII members
($-130<v_r<-90\kms$; the dashed lines in Figure~7) and assume the group with a higher velocity ($v_r\sim-50\kms$) to be
made of Galactic contaminants.

Ursa Major~II is the only observed satellite that does not show a clear radial velocity peak in our survey, even though
it is not the smallest of our samples. However, applying equation (1) to the probable UMaII members whatever their
radial velocity uncertainty yields a dispersion of only $7.4_{-2.8}^{+4.5}\kms$ around the systemic velocity
$-115\pm5\kms$ which is not unlike what is measured in brighter dwarf galaxies such as Bo\"otes. Interestingly, these
values are close to the expected value of  \citet{fellhauer06} who have tested the hypothesis of UMaII being the
progenitor of the Orphan Stream. They predict $v_r\sim-100\kms$ for UMaII and the rather high velocity dispersion we
measure seem to favor their scenario of a single-component satellite of $\sim10^5\msun$ that is on the verge of complete
disruption. Unfortunately, the DEIMOS sample does not extend over a wide enough range in declination to confirm the
$\sim10\kms$ gradient they predict along this direction.

Even though UMaII seems to be the progenitor of the Orphan Stream, the radial velocity measurement rules out a direct
link between UMaII and the H\textsc{i} velocity cloud Complex~A. Indeed, according to \citet{wakker91}, the cloud has a
radial velocity of $-160\kms$ whereas UMaII stars have $v_r=-115\kms$. This is not unexpected since Complex~A has a
distance that is lower than $10.1\kpc$, much lower than that of UMaII \citep{zucker06c}.

\section{Willman 1}

The peculiar object Willman~1 (Wil1, SDSS J1049+5103) was discovered as a very faint overdensity of
stars by \citet{willman05a} and later studied in more depth by \citet{willman06}. Having characteristics between that of
a globular cluster and an extremely faint dwarf galaxy, this satellite of the Milky Way has an absolute magnitude of
$M_V\sim-2.5$, a half-light radius $r_h=21\pm7\pc$ ($r_h=1.9''$) and resides at a distance of $38\pm7\kpc$.  Deep
observations reaching a magnitude $r\sim23.5$ suggest the object may be surrounded by multiple tidal tails.

\subsection{Photometry}
\begin{figure}
\begin{center}
\includegraphics[angle=270,width=\hsize]{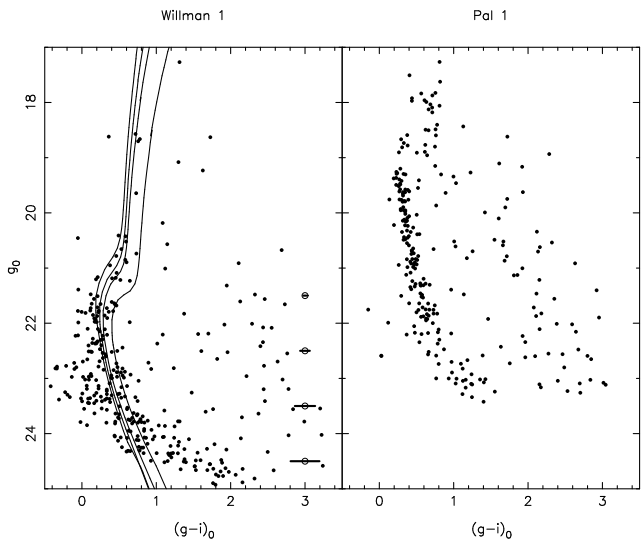}
\caption{Colour-magnitude diagram of the region within $5'$ of Willman~1 (left) and the equally faint globular cluster
Pal~1 (right) in the INT/WFC data. Although in both cases the main sequence is clearly visible, that of Wil1 is broader
than the single age/metallicity main sequence of Pal~1. This broadness cannot be explained by the photometric
uncertainties reported on the right of the Wil1 panel. The \citet{girardi04} isochrones with an age of 14\,Gyr and
metallicities of $\FeH=-2.3$, $-1.7$, $-1.3$ and $-0.7$ from left to right have also been overlaid on the CMD. A
metallicity spread covering this range could explain the broad main sequence.}
\end{center}
\end{figure}

Before analyzing the DEIMOS field that was targeted on Willman~1, we first present photometric data of the satellite
obtained using the Wide-Field Camera (WFC) mounted on the Isaac Newton Telescope (INT). One WFC pointing was observed in
the V and I band with integrations of 900\,s in both bands during the night of 20th November 2004 in photometric
conditions with seeing of $\sim1.0''$. Reduction was performed using the version of the CASU pipeline adapted for WFC
(see \citealt{irwin01} for more details). To provide a comparison with a globular cluster of similar luminosity, Pal~1
was also observed during the same night and under the same conditions except for a lower exposure time (600\,s for each
filter). The resulting CMDs are shown on Figure~8 with Wil1 in the left panel and Pal1 on the right panel. The
magnitudes of individual stars are de-reddened using the \citet{schlegel98} maps and transformed to SDSS magnitudes with
the color equations presented in \citet{ibata07}. Although both objects have a similar absolute magnitude, there is a
clear difference in the morphology of the main sequence of the two objects with the one of Wil1 being broader (at
$g\sim22.0$) than the main sequence of the globular cluster which harbours a single metallicity/age population.
Photometric uncertainties (reported on the right of the Wil1 panel) are too low to explain the spread of the Wil1 main
sequence in color. On the other hand, the color difference of the \citet{girardi04} isochrones for an old population
with metallicities ranging from $\FeH=-2.3$ to $\FeH=-0.7$ indicates that such a metallicity spread in the satellite
could explain the broadness of the main sequence observed in the satellite. This simple side by side comparison already
shows that Wil1 is not a simple cluster.

\begin{figure}
\begin{center}
\includegraphics[angle=270,width=\hsize]{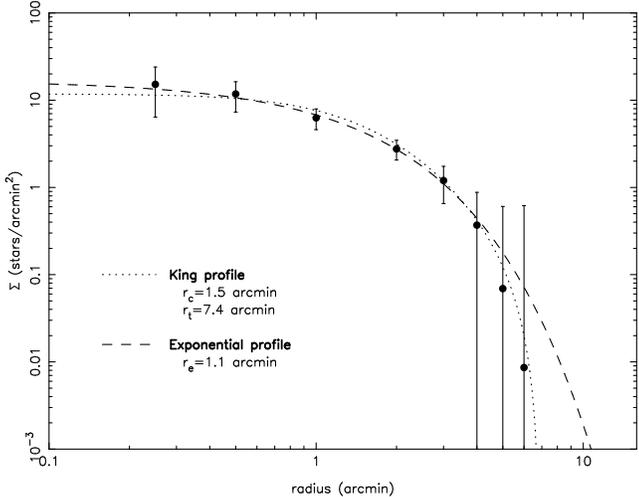}
\caption{Radial density profile of Willman~1 from the WFC data (filled circles) with error bars representing the
$1\sigma$ uncertainties, including background subtraction. The dotted line represents the best King profile that can
be fitted to the data within a radius of $4'$ while the dashed line represent the best exponential model. The
respective parameters (core radius, tidal radius and exponential length) are listed in the panel.}
\end{center}
\end{figure}

We use stars within the main sequence of Wil1 to draw the radial profile of the object. The possible tidal tails of the
object \citep{willman06} that also appear in our data and the low number of stars prevent the construction of a detailed
profile. Therefore, we simply construct the profile by determining the number of stars that are located within circular
annuli at increasing distance from the center of the object, taken as
$(\alpha,\delta)=(10\mathrm{h}49\mathrm{m}22\mathrm{s},+51\deg03'03\scnd6)$ (J2000; \citealt{willman05a}). Background
correction is determined from the number of stars within the selection box that are observed on the furthermost CCD of
the detector from the one containing Wil1 ($\sim1/3\deg$ away and covering $22.8\times11.4$\,arcmin$^2$). The resulting
profile appears in Figure~9 with the best fits for a King model (core radius $r_c=1\mcnd5$ and tidal radius
$r_t=7\mcnd4$) and an exponential model (exponential length $r_e=1\mcnd1$). Though the fits are admittedly not perfect,
the derived parameters can be used as rough estimates of the satellite structural parameters.

\subsection{Spectroscopy}
As before, the observed targets are shown on Figure~1\emph{e}, the SDSS CMD of this region of the sky on
Figure~2\emph{e} and the Na\textsc{i} doublet discrimination on Figure~3\emph{e}. The usual cut at
$\Sigma\textrm{Na}=0.8\Aa$ is very useful to isolate Wil1 stars from most of the foreground contaminants. For
homogeneity with the observations of the other satellites, the targets were selected from SDSS instead of the WFC
observations. Given the depth reached by DEIMOS observations ($i\sim22$), the extra depth provided by the WFC data is
not needed for target selection purposes. The observed stars and their parameters are listed in Table~5.

\begin{figure*}
\begin{center}
\includegraphics[angle=270,width=0.8\hsize]{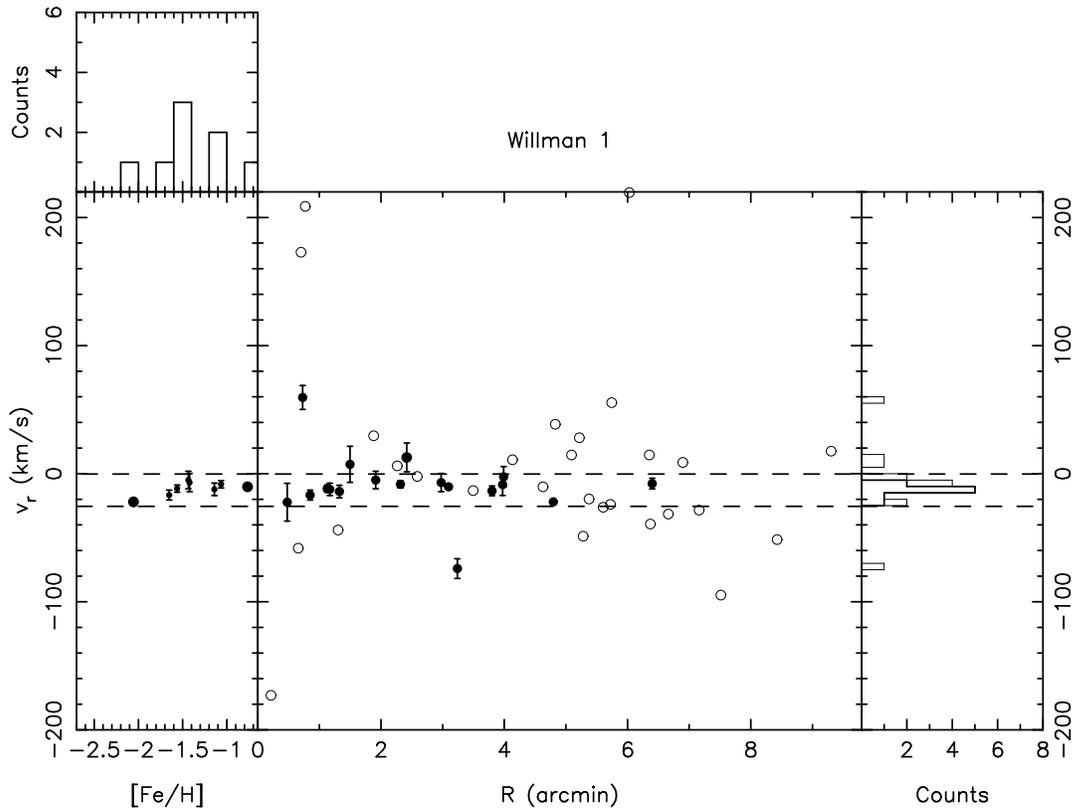}
\caption{Same as Figure~4 for the Willman~1 dwarf galaxy.}
\end{center}
\end{figure*}

The velocity distribution of the observed stars is shown on Figure~10 and reveals Wil1 stars are grouped in a low
dispersion peak that is well defined even though the satellite stars (filled circles in the central panel) overlap with
Galactic contaminants that have a higher dispersion (hollow circles). It is probable that some of the likely Wil1
members, selected from their location in the CMD are in fact contaminants but the low dispersion of the peak argues
against a significant fraction of such objects. We however refrain from assigning to the Wil1 population \emph{a priori}
field stars with the radial velocity of the peak and that are close to this RGB as we did in the case of Boo.

As for the Bo\"otes sample, we determine the parameters of the satellite by iteratively performing a $3\sigma$ clipping
and applying the same maximum likelihood technique, starting with the whole sample without the two clear outliers that
appear on Figure~10 at $v_r=59.5\kms$ and $v_r=-74.1\kms$. Convergence is reached for a systemic velocity of
$v_r=-12.3\pm2.5\kms$ ($v_{\mathrm{gsr}}=32.4\pm2.5\kms$) and a velocity dispersion of only
$\sigma=4.3_{-1.3}^{+2.3}\kms$ and higher than $2.1\kms$ at the 99\% confidence limit which means that the observations
are of a good enough quality to resolve the intrinsic dispersion of the population. Restricting the sample to the best
determined objects, with $v_{err}<6\kms$, yields very similar values ($v_r=-13.3\pm2.5\kms$ and
$\sigma=4.3_{-1.4}^{+2.6}\kms$).

Given the low velocity dispersion measured, one might get worried about the influence of binaries on the measured
dispersion. The work of \citet{olszewski96} shows that the effect of binaries at the center of Wil1 should be of
the order of $1-2\kms$. This value should be quadratically removed from the measured dispersion reducing $\sigma$
only marginally to $\sim3.8\kms$. The jitter that is known to exist in the atmosphere of red giant stars can also
artificially increase the measured velocity dispersion. In the Wil1 case however, all stars but two are more than 3
magnitudes fainter than the tip of the red giant branch whereas, according to \citet{cote96}, this is the magnitude
limit at which the jitter disappears. Therefore, this should not be an issue here.

The low velocity dispersion that is measured in Wil1 is not unlike what is found in some globular clusters
\citep[e.g.][]{dubath97} but the results from CVnI show that (faint) dwarf galaxies can also host such low velocity
dispersion populations. Moreover globular clusters host single metallicity and age populations and the
metallicity distribution of probable Wil1 members (left panels of Figure~10) show a wide spread in metallicity. Indeed,
the 8~stars with $S/N>10$ are scattered over $-2.0\lta \FeH \lta -1.0$. Given the low dispersion of the radial velocity
peak, it is also very unlikely that many of these stars are not Wil1 members. Alternatively, this large scatter could be
due to our extrapolating the determination of $\FeH$ from the Ca\textsc{ii} triplet equivalent width down to the
main-sequence turn-off or to the lower quality of the spectrum of most of these stars ($15>S/N>10$, smaller filled
circles). However, there is no sign of a correlation between the metallicities or the metallicity spread with the
magnitude of the stars whereas the isochrones overlaid on the CMD of Wil1 (Figure~10) and ongoing statistical analysis
of the SDSS CMD of this satellite reveals a similar metallicity spread (de Jong et al., in preparation). The two giant
stars that have $S/N>15$ are also spread over the whole metallicity range of the satellite (bigger filled circles in
the bottom left panel).

Hence, while the low velocity dispersion of Willman~1 is not unlike that observed in globular clusters, the large
metallicity spread argues strongly against such a possibility since globular clusters contain populations of a single age and
metallicity. This confirms the difference that is readily visible between the main sequences of Wil1 and Pal~1 in
Figure~8. Thus we conclude that even though Wil1 is very faint and small, it is probably a dwarf galaxy. The
metallicity spread could also explain the broad main sequence that is observed in Willman~1 (Figure~8).

As we have been doing for the other satellites, we estimate the mass of Wil1 from its central velocity dispersion and
structural parameters assuming it is in virial equilibrium. Although this is certainly not the case given the disturbed appearance
of the outskirts of the satellite as observed by \citet{willman06}, we will nevertheless assume the following techniques
provide a measure of its instantaneous mass. Equation (2) with $S_0=0.4\lsun\pc^{-2}$ from the exponential profile
fitted to the radial profile of Wil1 on Figure~8 and $r_{hb}=21\pc$ leads to $(M/L)\sim700$ in solar
units (i.e. $M\sim6\times10^5\msun$). Equation (3), using the core radius measured in \S~7.1, $r_c=16.6\pc$, yields 
$M=4.0\times10^5\msun$ (using $r_hb$ instead of $r_c$ as was done for the other dwarfs increases this
value to $M=5.2\times10^5\msun$). With a magnitude $M_V\sim-2.5$, Wil1 has a total luminosity $L\sim855\lsun$ which
yields $(M/L)\sim470$ in solar units. The two mass estimates agree although neither account for uncertainties in the
total luminosity of the satellite, which is as yet not well constrained. They show that Wil1 seems to be highly dark
matter dominated\footnote{using $\sigma\sim3.8\kms$ to account for binaries reduces the mass estimates by $\sim25\%$ but
does not change the conclusion that Wil1 is a dark matter dominated satellite.}, but with a mass that is only of the
order of $\sim5\times10^5\msun$.

\section{Discussion and conclusion}
The spectroscopic observations we have performed with the DEIMOS instrument on new faint Milky Way satellites reveal
that these objects are diverse and more complex than could be expected from their photometry alone (see Table~6 for a
summary). In particular, UMaII does not show any clear radial velocity signal and comparison with the
\citet{fellhauer06} simulations favors a direct link with the Orphan Stream. The other four satellites have radial
velocity dispersions that are consistent with dwarf galaxies, although this does not mean they are comparable. In
particular, CVnI presents a complex structure with two very distinct populations: more metal-rich stars forming a
kinematically cold population at the center of the satellite and more metal-poor stars that belong to a more extended
and hotter population \citep{ibata06}. A similar behaviour could also be present in UMaI though our sample remains too
small to derive reliable parameters for this satellite. On the other hand, Boo and Wil1 show much cleaner radial
velocity peaks.

Overall, if these objects are indeed dwarf galaxies, their median metallicity is never lower than
$\FeH\sim-2.1$ (within the metallicity scale caveat mentioned in \S~3). Therefore, the $M_V-\FeH$ relation that is found
for brighter galaxies (see for instance \citealt{mateo98} or Fig.~9 of \citealt{martin06}) seem to break down or at
least spread out at faint magnitudes. Following the previous relation, one would expect that satellites such as Bo\"otes
have a metallicity of $\FeH\sim-2.5$ and even lower for Wil1. Moreover in Boo and CVnI for which the samples are the
more populated and the more reliable (since only giant or sub-giant stars were targeted), there still is a dearth of
very metal-poor stars, as in the brighter dwarf galaxies Sculptor, Sextans, Fornax and Carina \citep{helmi06}.

Similarly to brighter satellites, these objects appear to be highly dark matter dominated when applying the
\citet{richstone86} or \citet{illingworth76} equations -- under the usual assumptions that the systems are spherical and
in virial equilibrium -- to estimate their mass-to-light ratio or mass from their central velocity dispersion. However, 
Willman~1 appears to be a peculiar outlier to the inferred mass distribution of faint satellites. Although it does
appear highly dark-matter dominated with a mass-to-light ratio estimate ranging from $\sim470$ to $\sim700$, this
object is at least one order of magnitude less massive ($M\sim5\times10^5\msun$) than brighter dwarf galaxies who share
masses of $\sim10^7\msun$ or higher.

\begin{figure}
\begin{center}
\includegraphics[width=\hsize]{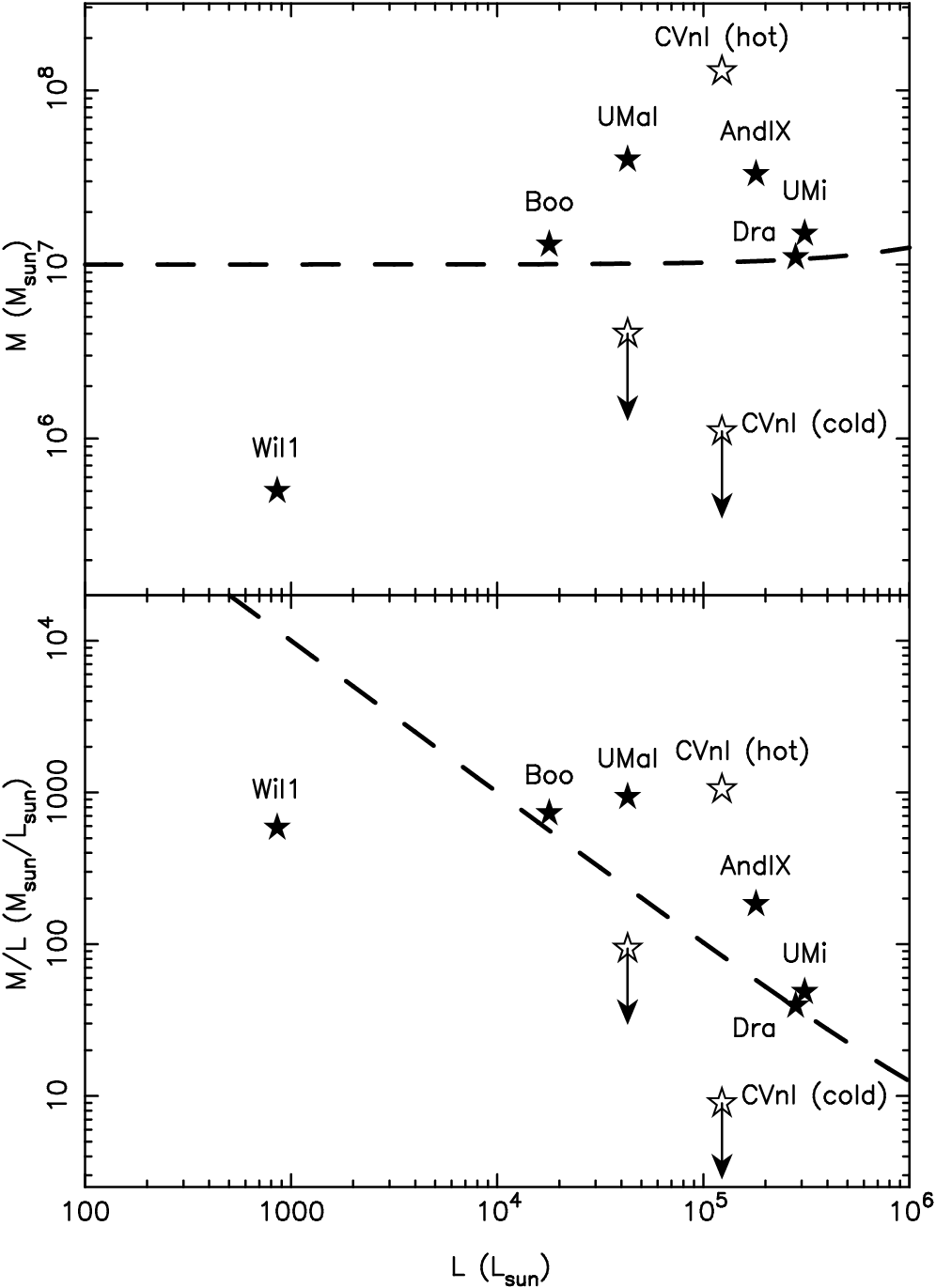}
\caption{Comparison of the mass-to-light ratios ($M/L$) and mass estimates ($M$) for the faintest known dwarf galaxies
with central velocity dispersion estimates (filled stars). In all cases, the mass estimates were derived using equation
(3). For Draco and Ursa Minor, we use the core radii and velocity dispersions quoted in \citet{irwin95} and
\citet{armandroff95} while we use those quoted in \citet{chapman05} for AndIX. For UMaI, the hollow star represent the
99\% confidence higher mass limit obtained from the cold component they it may harbor (see \S~5) while for CVnI, the two
hollow stars represent the two diverging estimates obtained using either the cold more metal-rich half of the sample or
the hot metal-poor half \citet{ibata06}.}
\end{center}
\end{figure}

With the mass we infer, it is readily visible in Figure~11 that Wil1 is also a clear outlier from the \citet{mateo98}
relation between $M/L$ and the luminosity of a dwarf galaxies in the Local Group. This is not surprising since this
relation ($M/L=2.5+ 10^7/(L/\lsun)$) naturally assumes that dwarf galaxies have masses higher than $10^7\msun$. Even
though the uncertainties are sizeable, Wil1 would need to be at least ten times more massive to follow the relation.
Keeping the same structural parameters, this would mean that the central velocity dispersion would have to be at least
$\sqrt{10}\sim3$ times higher than the one we have measured, that is $\sim13\kms$, which is hardly compatible with the
DEIMOS observations. Does it mean that a numerous population of small satellites (such as the structurally similar
Segue~1; \citealt{belokurov06c}) residing in less massive dark matter halos than brighter dwarf galaxies
($\lta10^7\msun$) have until now eluded us? Or does it mean that Wil1 was once a more luminous and massive dwarf galaxy
and that its outskirts have been stripped out by tidal interaction with the Milky Way, leaving only its central
population visible at present? 

Firstly, it has to be noted that equation (2) and (3) are only valid for a system with a constant mass-to-light ratio.
The mass estimates determined here only correspond to the central regions of the satellites where stars can be used
as tracers. Therefore, it does not rule out \textit{per se} that Wil1 could be embedded in a dark matter halo that is
as massive as brighter galaxies. However, the low central mass estimate of this dark matter halo compared to the brighter
galaxies would still hint at a halo with a lower central density and still make Wil1 a peculiar object.

The heating of an initially colder central population could also produce the observed velocity dispersion. While the
metallicity spread that is measured within Wil1 argues against the simple disruption of a globular cluster, one could
argue that Wil1 is the remnant of a CVnI-like structure with only the colder core population remaining after stripping of
the hotter component by tidal interaction with the Milky Way. However, in such a scenario, a significant amount of
material would have to have been stripped to explain the absence of an underlying hot component in the spectroscopic
data, at odds with the photometric data that do not show tidal tails containing a significant part of the whole
satellite \citep{willman06}. Moreover, according to \citet{piatek95} the influence of tidal interaction of the
measured velocity dispersion becomes significant at high distance from the center of the satellite whereas for the
Wil1 sample that extends to $\sim2r_{hb}$, there is no visible increase  of the dispersion with distance in Figure~10.
Although the number of stars is too low for a detailed analysis, this does not favor the presence of strong tidal tails.
Thus it would seem more likely that Wil1 is a highly dark matter dominated object although it resides in a much less
massive dark matter halo than those of brighter dwarf galaxies such as Boo. The small systemic velocity of this peculiar
object ($v_{\mathrm{gsr}}=33.0\kms$) could mean it does not have a strongly radial orbit around the Milky Way, which
could in turn explain why this object has survived until now. 

Though dark matter halos are expected to form down to planet-mass structures \citep{diemand05}, a minimum mass
is required to have a deep-enough potential to retain gas and eventually form stars. Therefore Wil1 could be of
significant help in understanding how the lowest-mass systems form if it is confirmed to inhabit a low mass dark matter halo. Such a
confirmation could come from the search for extra-tidal stars whose presence or absence would make it clearer if it
is an unbound alignment of stars, a surviving core or a complete system. Besides, the presence of kinematically
different populations in CVnI \citep{ibata06} and perhaps UMaI yields significantly different mass estimates. Plotting
these mass estimates on Figure~11 (hollow stars) yields a significant spread in the mass-luminosity relation that
should be taken as a warning against using $M/L$ ratios when precise structural parameters are unknown. Moreover, it
cannot be excluded at the moment that these satellites could have been strongly disrupted and influenced by tidal
interaction or that they may have recently accreted some stellar material that did not have the time to relax in the
gravitational potential of the dwarf. If this is the case, the simple application of equations (2) and (3) are not
warranted and would lead to erroneous mass estimates.

The promising five new faint satellites presented by \citet{belokurov06c} from the analysis of the latest data release of
the SDSS, with luminosities in the range $1.3\times10^3\lta L \lta2\times10^4\lsun$ will also undoubtedly help us
understand the low-luminosity regime of dwarf galaxies between Wil1 and Boo. Spectroscopic observations for these
objects are crucial to confirm if the $\sim10^7\msun$ limit still holds or if there is indeed a population of very faint,
but not completely dark galaxies inhabiting less massive dark matter halos that orbit within the Local Group.

\section*{Acknowledgments}
We would like to acknowledge the referee, Steven Majewski, for a thorough reading of the paper and comments that
improved its quality. NFM would like to thank Vasily Belokurov, Mike Fellhauer and Dan Zucker for fruitful discussions
at the ``Dissecting the Milky Way'' Workshop that took place in Leiden in November 2006 as well as Amina Helmi and
Hans-Walter Rix for organising such an interesting and exciting workshop.

\newcommand{\mnras}{MNRAS}
\newcommand{\pasa}{PASA}
\newcommand{\nat}{Nature}
\newcommand{\araa}{ARAA}
\newcommand{\aj}{AJ}
\newcommand{\apj}{ApJ}
\newcommand{\apjl}{ApJ}
\newcommand{\apjs}{ApJSupp}
\newcommand{\aap}{A\&A}
\newcommand{\aaps}{A\&ASupp}
\newcommand{\pasp}{PASP}

\begin{table*}
\begin{center}
\caption{Derived parameters for stars in the Bo\"otes sample.}
\label{tableSat}
\begin{tabular}{llccccccccc}
\hline
$\alpha$ (J2000) & $\delta$ (J2000) & g & i & $v_r$ ($\kms$) & $v_{err}$ ($\kms$) & $S/N$ & $\FeH$  &
$\Sigma\textrm{Ca}$ & $\Sigma\textrm{Na}$ & member?\\\hline
14 00 23.75 &  14 30 46.3 & 18.64 & 16.45 &  -45.7 &   1.0 &  73 &  --  &  3.6 &  2.2 & n\\
13 59 41.24 &  14 33 09.4 & 17.98 & 17.39 &  -64.6 &   0.8 &  63 &  --  &  2.9 &  0.5 & n\\
14 00 31.82 &  14 32 24.4 & 17.74 & 17.29 &    7.3 &   1.2 &  80 &  --  &  2.4 &  0.4 & n\\
14 00 10.49 &  14 31 45.5 & 18.19 & 17.13 &   98.2 &   0.6 &  54 & -2.2 &  2.3 &  0.4 & y\\
14 00 12.92 &  14 33 11.8 & 19.54 & 18.72 &  100.1 &   0.9 &  41 & -1.9 &  2.0 &  0.5 & y\\
14 00 33.08 &  14 29 59.7 & 19.74 & 18.92 &   95.6 &   0.9 &  43 & -2.2 &  1.3 &  0.3 & y\\
14 00 22.97 &  14 30 45.0 & 18.55 & 17.50 &  -45.0 &   1.4 &  57 &  --  &  4.6 &  1.2 & n\\
14 00 03.08 &  14 30 23.6 & 21.45 & 20.78 &   97.9 &   4.5 &  10 & -1.3 &  2.3 &  0.0 & y\\
14 00 24.62 &  14 31 59.6 & 21.33 & 20.73 &  -74.3 &   3.3 &  11 &  --  &  2.8 &  0.1 & n\\
13 59 44.27 &  14 32 41.2 & 21.28 & 20.61 &   99.7 &  13.0 &   8 &  --  &  1.7 &  0.3 & y\\
13 59 52.33 &  14 32 45.7 & 20.67 & 19.95 &   97.9 &   1.5 &  19 & -1.6 &  2.1 &  0.4 & y\\
13 59 53.76 &  14 30 56.0 & 20.65 & 19.88 &  119.6 &   1.9 &  21 &  --  &  3.8 &  0.2 & n\\
14 00 05.34 &  14 30 23.3 & 21.04 & 20.41 &   95.4 &   2.5 &  11 & -1.5 &  2.1 &  0.1 & y\\
14 00 00.83 &  14 33 07.7 & 19.06 & 18.04 &   32.6 &   1.2 &  35 &  --  &  4.4 &  0.5 & n\\
13 59 44.96 &  14 32 30.1 & 20.77 & 20.07 &   98.0 &   1.8 &  18 & -1.6 &  2.1 &  0.3 & y\\
14 00 15.42 &  14 32 08.5 & 22.28 & 21.69 & -235.0 &  13.6 &   5 &  --  &  9.2 &  0.9 & n\\
14 00 23.38 &  14 32 45.3 & 21.67 & 21.12 &  102.2 &  12.1 &   6 &  --  &  1.0 &  0.3 & y\\
13 59 51.08 &  14 30 49.8 & 21.96 & 21.30 &   94.0 &   3.2 &   5 &  --  &  1.7 &  0.2 & y\\
13 59 50.76 &  14 31 14.2 & 21.75 & 21.01 &   85.2 &   5.7 &   8 &  --  &  1.5 &  0.1 & y\\
13 59 48.85 &  14 33 43.6 & 22.09 & 21.57 & -443.4 &   6.4 &   3 &  --  &  0.6 &  0.0 & n\\
14 00 12.25 &  14 31 51.9 & 18.52 & 18.05 &   -5.4 &   0.6 &  57 &  --  &  1.7 &  0.5 & n\\
14 00 13.93 &  14 33 04.4 & 20.43 & 18.16 &   -6.5 &   0.9 &  52 &  --  &  3.1 &  1.6 & n\\
14 00 17.14 &  14 33 48.6 & 19.86 & 19.44 &  -43.0 &   1.2 &  26 &  --  &  1.8 &  0.0 & n\\
14 00 24.94 &  14 30 56.9 & 20.40 & 19.34 &  -55.5 &   1.6 &  27 &  --  &  4.7 &  0.8 & n\\
14 00 27.29 &  14 32 19.6 & 20.31 & 19.66 &  103.9 &   1.5 &  25 & -1.9 &  1.5 &  0.3 & y\\
14 00 22.45 &  14 33 26.9 & 19.54 & 19.00 &   95.8 &   1.7 &  29 & -2.1 &  1.5 &  0.2 & y\\
14 00 25.71 &  14 34 15.4 & 20.56 & 18.31 &   35.5 &   1.1 &  48 &  --  &  3.3 &  0.0 & n\\
14 00 06.98 &  14 32 07.8 & 18.55 & 17.74 &   37.5 &   1.1 &  57 &  --  &  4.2 &  1.0 & n\\
14 00 28.73 &  14 31 18.2 & 20.78 & 18.34 &   16.2 &   1.0 &  51 &  --  &  2.8 &  2.5 & n\\
14 00 07.35 &  14 31 51.1 & 20.61 & 18.16 &  -17.7 &   0.9 &  49 &  --  &  2.8 &  2.8 & n\\
13 59 36.02 &  14 34 07.1 & 18.38 & 17.98 &   54.1 &   1.4 &  53 &  --  &  2.8 &  0.0 & n\\
13 59 40.86 &  14 32 48.3 & 21.71 & 18.98 &  -50.2 &   1.5 &  32 &  --  &  1.7 &  3.2 & n\\
14 00 29.72 &  14 31 50.0 & 20.16 & 19.55 &   32.6 &   1.9 &  24 &  --  &  3.6 &  0.3 & n\\
13 59 45.95 &  14 31 40.6 & 19.91 & 19.65 &  131.3 &  12.3 &  24 &  --  &  4.7 &  0.4 & n\\
13 59 48.34 &  14 32 03.6 & 19.65 & 19.24 &  101.9 &   1.2 &  30 & -2.1 &  1.6 &  0.5 & y\\
13 59 40.29 &  14 30 59.7 & 21.05 & 18.94 &   -4.8 &   1.5 &  36 &  --  &  3.4 &  1.8 & n\\
13 59 31.58 &  14 32 55.3 & 20.74 & 19.14 &   14.6 &   1.3 &  33 &  --  &  3.7 &  1.3 & n\\
13 59 33.58 &  14 30 44.6 & 19.80 & 19.33 &    3.2 &   1.2 &  30 &  --  &  3.0 &  0.4 & n\\
14 00 05.67 &  14 33 29.8 & 21.64 & 19.66 &   15.9 &   2.5 &  25 &  --  &  3.4 &  0.0 & n\\
14 00 08.99 &  14 33 26.6 & 21.38 & 19.32 &  -12.2 &   2.0 &  33 &  --  &  3.3 &  0.0 & n\\
14 00 09.89 &  14 30 52.0 & 20.89 & 19.03 &  -16.5 &   1.2 &  35 &  --  &  3.8 &  1.5 & n\\
14 00 04.87 &  14 31 44.2 & 21.78 & 19.92 &   12.0 &   4.4 &  15 &  --  &  3.3 &  1.2 & n\\
13 59 39.84 &  14 34 33.4 & 19.42 & 18.72 &    0.8 &   1.1 &  36 &  --  &  1.7 &  0.0 & n\\
13 59 56.61 &  14 32 43.3 & 21.76 & 19.24 &  -43.0 &   2.4 &  32 &  --  &  2.5 &  2.8 & n\\
13 59 46.27 &  14 34 09.2 & 19.84 & 19.18 &  136.6 &   1.7 &  23 &  --  &  2.2 &  0.0 & n\\
13 59 47.72 &  14 30 50.6 & 20.52 & 18.42 &  -22.9 &   1.0 &  46 &  --  &  3.3 &  1.9 & n\\
13 59 48.04 &  14 31 16.3 & 20.30 & 18.14 &  -17.6 &   1.2 &  46 &  --  &  3.5 &  2.3 & n\\
14 00 32.48 &  14 30 29.9 & 22.60 & 19.66 &  -13.9 &   1.4 &  29 &  --  &  1.5 &  3.4 & n\\
14 00 04.61 &  14 31 28.1 & 22.39 & 19.45 &  -27.5 &   1.2 &  32 &  --  &  1.9 &  3.2 & n\\
13 59 57.69 &  14 29 55.4 & 17.76 & 14.84 &  -23.5 &   2.0 &  48 &  --  &  1.8 &  3.6 & n\\
13 59 43.83 &  14 32 48.8 & 20.86 & 18.96 &  -44.7 &   1.2 &  40 &  --  &  3.6 &  1.3 & n\\
14 00 18.66 &  14 33 23.1 & 18.30 & 17.76 &   10.9 &   0.9 &  60 &  --  &  3.5 &  0.7 & n\\
\hline
\end{tabular}
\end{center}
Columns (1) and (2) correspond to the J2000 coordinates of the stars as they appear in SDSS. Columns (3) and (4) are 
he g and i magnitudes observed by the SDSS. Columns (5) and (6) are the measured radial velocity and the corresponding
uncertainty. Column (7) is the signal-over-noise ratio of the spectrum, column (8) is the derived metallicity of Bo\"otes
members with $S/N>10$ per \AA, column (9) is the weighted sum of the equivalent widths of the three Ca\textsc{i} lines at
8498, 8542 and $8662\Aa$, column (10) is the sum of the equivalent widths of the two Na\textsc{i} lines at 8193 and
$8195\Aa$ and column (11) has the value 'y' for stars that were used to derive the parameters of the dwarf.
None members ('n' value) either have $v_{err}>15\kms$, are not aligned with the Bo\"otes CMD features, have $\FeH>-1.0$,
have $\Sigma\textrm{Na}>0.8$ or are clearly away from the radial velocity peak produced by the dwarf galaxy.
\end{table*}

\begin{table*}
\begin{center}
\addtocounter{table}{-1}
\caption{\emph{continued}}
\label{tableSat}
\begin{tabular}{llccccccccc}
\hline
$\alpha$ (J2000) & $\delta$ (J2000) & g & i & $v_r$ ($\kms$) & $v_{err}$ ($\kms$) & $S/N$ & $\FeH$  &
$\Sigma\textrm{Ca}$ & $\Sigma\textrm{Na}$ & member?\\\hline
13 59 22.71 &  14 25 56.8 & 15.37 & 14.32 &  153.6 &  12.4 & 151 &  --  &  3.9 &  1.2 & n\\
14 00 20.11 &  14 29 23.6 & 19.49 & 17.17 &   67.1 &   1.7 &  76 &  --  &  2.9 &  1.6 & n\\
13 59 33.85 &  14 29 12.1 & 18.07 & 17.40 &    9.9 &   1.2 &  61 &  --  &  3.8 &  0.4 & n\\
14 00 19.77 &  14 27 04.6 & 19.60 & 17.11 &   13.0 &   1.1 &  70 &  --  &  2.6 &  2.9 & n\\
13 59 57.85 &  14 28 02.5 & 20.40 & 19.72 &  102.4 &   4.4 &  18 & -2.0 &  1.3 &  0.4 & y\\
13 59 39.37 &  14 26 38.4 & 20.38 & 19.67 &   97.6 &   7.2 &  21 & -2.2 &  0.9 &  0.3 & y\\
13 59 42.19 &  14 29 42.3 & 19.62 & 18.77 &   86.6 &   1.8 &  37 & -2.3 &  1.2 &  0.0 & y\\
14 00 03.33 &  14 28 51.5 & 20.76 & 20.03 &  115.9 &   3.0 &  18 & -1.9 &  1.3 &  0.3 & y\\
14 00 05.61 &  14 26 18.9 & 20.43 & 19.71 &  108.1 &   1.5 &  22 & -2.0 &  1.3 &  0.5 & y\\
13 59 47.07 &  14 28 52.6 & 21.48 & 20.92 &  101.6 &   5.5 &   8 &  --  &  1.7 &  0.0 & y\\
14 00 09.85 &  14 28 23.0 & 18.45 & 17.39 &   92.6 &   0.6 &  60 & -2.7 &  1.1 &  0.5 & y\\
13 59 50.63 &  14 29 11.1 & 19.12 & 18.23 &   92.8 &   1.5 &  56 & -2.3 &  1.5 &  0.0 & y\\
14 00 11.54 &  14 25 56.1 & 21.54 & 20.88 &  108.7 &   4.2 &   9 &  --  &  1.4 &  0.3 & y\\
14 00 25.83 &  14 26 07.6 & 18.38 & 17.37 &  104.9 &   0.8 &  48 & -1.6 &  3.5 &  0.6 & y\\
14 00 13.84 &  14 26 07.9 & 20.56 & 19.85 &  188.6 &   6.1 &  14 &  --  &  1.0 &  0.2 & n\\
14 00 21.84 &  14 25 53.4 & 21.00 & 20.33 &  112.1 &   1.5 &  14 & -1.9 &  1.2 &  0.4 & y\\
14 00 26.18 &  14 27 29.5 & 19.08 & 18.24 &   70.4 &   1.8 &  41 &  --  &  3.7 &  0.8 & n\\
13 59 33.51 &  14 28 21.6 & 21.91 & 21.23 &   84.4 &   6.6 &   6 &  --  &  2.1 &  0.2 & y\\
13 59 34.38 &  14 30 17.2 & 22.02 & 21.29 &   86.0 &  14.8 &   5 &  --  &  0.3 &  0.0 & y\\
13 59 32.07 &  14 27 12.5 & 22.46 & 21.91 & -416.3 &  11.8 &   4 &  --  &  0.6 &  2.1 & n\\
13 59 32.74 &  14 28 23.5 & 22.05 & 21.50 & -122.5 &   4.1 &   5 &  --  &  2.2 &  0.1 & n\\
14 00 02.29 &  14 26 53.4 & 21.97 & 21.38 &  104.0 &   9.7 &   7 &  --  &  1.2 &  0.2 & y\\
13 59 45.71 &  14 25 52.6 & 22.25 & 21.56 &   93.8 &   4.5 &   5 &  --  & 12.4 &  0.2 & y\\
14 00 01.73 &  14 26 16.2 & 22.29 & 21.82 &  239.4 &  10.2 &   4 &  --  &  2.3 &  0.4 & n\\
14 00 28.70 &  14 27 05.2 & 21.71 & 21.13 &  209.3 &  10.9 &   6 &  --  &  1.4 &  0.1 & n\\
14 00 23.34 &  14 26 08.0 & 21.87 & 21.22 &   99.9 &   5.4 &   7 &  --  &  2.2 &  0.3 & y\\
14 00 25.04 &  14 26 27.0 & 22.11 & 21.53 &  170.4 &   9.2 &   5 &  --  &  0.1 &  0.1 & n\\
14 00 22.78 &  14 29 20.5 & 18.51 & 18.05 &   30.2 &   5.3 &  52 &  --  &  2.2 &  0.6 & n\\
14 00 15.97 &  14 30 24.8 & 18.05 & 17.69 & -147.4 &   1.7 &  65 &  --  &  1.5 &  0.0 & n\\
13 59 36.00 &  14 29 38.5 & 19.04 & 18.57 &  -50.1 &   1.2 &  39 &  --  &  2.6 &  0.0 & n\\
13 59 26.54 &  14 26 45.3 & 19.08 & 18.64 &   25.8 &   2.0 &  33 &  --  &  2.3 &  0.5 & n\\
13 59 26.24 &  14 29 29.2 & 22.56 & 19.88 &  194.1 &   4.2 &  12 &  --  &  3.0 &  3.0 & n\\
13 59 37.39 &  14 29 41.6 & 20.94 & 18.93 &  147.1 &   1.9 &  36 &  --  &  3.7 &  0.0 & n\\
13 59 45.17 &  14 29 03.9 & 20.14 & 17.67 &   -1.7 &   1.9 &  55 &  --  &  2.7 &  1.7 & n\\
14 00 04.12 &  14 27 36.7 & 19.74 & 19.32 &    3.1 &   2.2 &  26 &  --  &  2.2 &  0.6 & n\\
14 00 04.62 &  14 26 08.4 & 21.66 & 19.26 &   -9.8 &   1.4 &  30 &  --  &  3.1 &  2.1 & n\\
13 59 49.62 &  14 28 12.0 & 20.24 & 19.80 &  -85.6 &   5.9 &  17 &  --  &  2.4 &  0.3 & n\\
14 00 10.26 &  14 26 00.4 & 19.86 & 19.26 &   -7.4 &   1.4 &  27 &  --  &  1.7 &  0.4 & n\\
13 59 43.92 &  14 29 00.0 & 21.43 & 18.93 &  -27.9 &   1.6 &  35 &  --  &  2.6 &  1.2 & n\\
13 59 52.16 &  14 28 37.7 & 21.00 & 18.68 &  -29.9 &   2.5 &  38 &  --  &  2.8 &  2.6 & n\\
13 59 50.91 &  14 28 42.6 & 20.20 & 19.88 &  182.2 &   9.8 &  16 &  --  & 44.4 &  0.3 & n\\
14 00 06.65 &  14 28 49.6 & 21.92 & 19.01 &   30.2 &   1.2 &  32 &  --  &  1.8 &  3.5 & n\\
14 00 14.85 &  14 29 09.6 & 19.55 & 19.07 &   -2.8 &   1.6 &  27 &  --  &  2.2 &  0.7 & n\\
14 00 15.18 &  14 27 49.8 & 17.98 & 15.67 &   20.2 &   1.9 &  60 &  --  &  3.0 &  2.9 & n\\
\hline
\end{tabular}
\end{center}
\end{table*}

\begin{table*}
\begin{center}
\caption{Derived parameters for stars in the Canes Venatici~I sample.}
\label{tableSat}
\begin{tabular}{llccccccccc}
\hline
$\alpha$ (J2000) & $\delta$ (J2000) & g & i & $v_r$ ($\kms$) & $v_{err}$ ($\kms$) & $S/N$ & $\FeH$  &
$\Sigma\textrm{Ca}$ & $\Sigma\textrm{Na}$ & member?\\\hline
13 27 33.02 &  33 37 54.5 & 17.26 & 16.57 &  -86.1 &   1.0 &  62 &  --  &  4.1 &  1.4 & n\\
13 28 25.01 &  33 34 58.9 & 19.77 & 17.27 &  -41.7 &   1.2 &  40 &  --  &  2.8 &  2.4 & n\\
13 28 41.21 &  33 36 34.2 & 19.76 & 16.80 &   -8.9 &   2.0 &  47 &  --  &  1.6 &  3.8 & n\\
13 27 27.51 &  33 36 25.0 & 18.27 & 16.51 &  -40.0 &   1.0 &  63 &  --  &  4.5 &  1.6 & n\\
13 27 36.41 &  33 37 16.8 & 19.76 & 18.29 &    7.5 &   0.6 &  56 & -2.3 &  2.9 &  0.5 & y\\
13 27 49.42 &  33 38 09.0 & 21.03 & 19.95 &   15.9 &   1.2 &  18 & -2.5 &  1.4 &  0.0 & y\\
13 28 13.74 &  33 35 55.6 & 20.20 & 18.93 &   20.7 &   0.8 &  32 & -2.0 &  3.1 &  0.3 & y\\
13 28 28.27 &  33 37 02.8 & 20.03 & 18.60 &   25.5 &   0.6 &  42 & -2.2 &  2.8 &  0.3 & y\\
13 28 27.56 &  33 36 43.2 & 19.92 & 18.42 &   30.6 &   0.6 &  42 & -2.3 &  2.7 &  0.3 & y\\
13 28 07.83 &  33 36 38.3 & 21.45 & 20.57 &   34.6 &   2.3 &  12 & -2.1 &  2.0 &  0.0 & y\\
13 27 58.61 &  33 36 54.0 & 20.20 & 18.84 &   22.3 &   0.7 &  40 & -1.8 &  3.8 &  0.5 & y\\
13 28 10.82 &  33 36 31.9 & 21.23 & 20.18 &   17.5 &   2.4 &   9 &  --  &  2.7 &  0.5 & y\\
13 28 20.95 &  33 38 04.8 & 20.14 & 18.74 &   23.6 &   0.7 &  41 & -1.9 &  3.5 &  0.0 & y\\
13 28 20.63 &  33 38 04.3 & 21.51 & 20.61 &   30.8 &   2.2 &  12 & -1.9 &  2.4 &  0.0 & y\\
13 28 21.84 &  33 38 41.0 & 20.14 & 18.72 &   34.8 &   0.6 &  49 & -2.2 &  2.7 &  0.0 & y\\
13 28 02.64 &  33 36 06.8 & 21.52 & 20.61 &   26.7 &   2.0 &  13 & -1.9 &  2.5 &  0.1 & y\\
13 28 29.71 &  33 36 28.6 & 21.52 & 20.50 &   22.5 &   2.2 &  11 & -2.0 &  2.2 &  0.4 & y\\
13 28 39.76 &  33 35 37.2 & 21.64 & 20.87 &   34.8 &   1.9 &  13 & -1.8 &  2.7 &  0.1 & y\\
13 28 17.67 &  33 35 11.0 & 21.76 & 20.81 &   25.4 &   3.7 &  12 & -1.5 &  3.3 &  0.0 & y\\
13 27 57.78 &  33 35 56.3 & 21.84 & 20.91 &   36.1 &   4.0 &   9 &  --  &  2.4 &  0.5 & y\\
13 27 49.74 &  33 34 09.5 & 21.55 & 20.63 &   20.9 &   2.2 &  11 & -1.8 &  2.8 &  0.3 & y\\
13 27 34.95 &  33 34 30.6 & 21.80 & 20.84 &   27.7 &   5.1 &  10 & -2.3 &  1.4 &  0.1 & y\\
13 27 30.41 &  33 35 44.3 & 21.67 & 20.78 &   21.7 &   2.7 &  13 & -1.8 &  2.7 &  0.1 & y\\
13 27 59.57 &  33 34 03.9 & 21.72 & 20.79 &   27.2 &   2.5 &  11 & -2.3 &  2.3 &  0.4 & y\\
13 28 16.06 &  33 34 53.2 & 21.02 & 19.95 &   36.7 &   1.1 &  23 & -2.1 &  2.5 &  0.2 & y\\
13 28 26.79 &  33 34 39.7 & 21.45 & 20.58 &   26.9 &   3.6 &  11 & -2.4 &  1.3 &  0.3 & y\\
13 28 11.71 &  33 34 07.8 & 21.82 & 20.84 &   27.4 &   3.4 &   9 &  --  &  2.6 &  0.1 & y\\
13 27 56.03 &  33 34 25.5 & 21.06 & 20.00 &   23.5 &   1.0 &  20 & -2.0 &  2.6 &  0.3 & y\\
13 27 34.01 &  33 36 23.8 & 21.19 & 20.08 &   20.8 &   1.4 &  17 & -1.9 &  2.7 &  0.2 & y\\
13 28 10.35 &  33 34 27.7 & 20.67 & 19.54 &   22.2 &   0.8 &  30 & -1.8 &  3.3 &  0.4 & y\\
13 27 51.59 &  33 35 43.2 & 21.16 & 20.03 &   14.4 &   2.5 &  17 & -2.0 &  2.5 &  0.2 & y\\
13 28 08.82 &  33 34 41.2 & 20.43 & 19.24 &   21.9 &   0.7 &  33 & -2.1 &  2.8 &  0.2 & y\\
13 27 47.82 &  33 36 54.2 & 22.49 & 21.80 &   34.2 &   4.0 &   5 &  --  &  2.6 &  0.1 & y\\
13 28 00.00 &  33 36 28.2 & 22.03 & 21.20 &   36.9 &   5.0 &   7 &  --  &  2.8 &  0.2 & y\\
13 28 04.85 &  33 35 58.5 & 22.03 & 21.34 &   44.3 &   4.5 &   9 &  --  &  2.3 &  0.4 & y\\
13 28 22.77 &  33 36 44.7 & 22.63 & 21.87 &   41.7 &   8.9 &   5 &  --  &  2.7 &  0.4 & y\\
13 28 38.38 &  33 34 25.7 & 22.13 & 21.32 &   11.9 &   3.6 &   7 &  --  &  1.7 &  0.2 & y\\
13 28 34.77 &  33 35 07.1 & 22.03 & 21.78 &  215.9 &  10.3 &   3 &  --  &  4.0 &  0.3 & n\\
13 28 16.48 &  33 37 16.9 & 22.44 & 21.59 &   36.8 &   3.4 &   6 &  --  &  2.4 &  0.7 & y\\
13 28 00.83 &  33 36 05.6 & 22.33 & 21.38 &    8.6 &   3.6 &   7 &  --  &  2.1 &  0.0 & y\\
13 27 57.16 &  33 36 34.4 & 22.53 & 21.77 &   11.8 &   6.5 &   5 &  --  & 61.1 &  0.5 & y\\
13 27 35.36 &  33 35 17.5 & 22.43 & 21.67 &   32.4 &   6.4 &   5 &  --  &  3.2 &  0.1 & y\\
13 27 37.37 &  33 35 55.6 & 22.29 & 21.98 &   38.3 &   3.2 &   4 &  --  &  6.4 &  0.2 & y\\
13 27 52.51 &  33 35 47.4 & 22.03 & 21.28 &  229.7 &   3.3 &   6 &  --  &  1.9 &  0.3 & n\\
13 28 05.88 &  33 34 12.0 & 22.24 & 21.50 &   27.1 &   2.8 &   8 &  --  &  3.3 &  0.1 & y\\
13 28 09.55 &  33 34 43.2 & 22.62 & 21.74 &   24.6 &   5.1 &   7 &  --  &  2.8 &  0.3 & y\\
13 28 08.50 &  33 34 26.4 & 22.83 & 21.97 &   19.9 &  10.3 &   4 &  --  &  5.6 &  0.8 & n\\
13 27 54.82 &  33 34 22.1 & 22.52 & 21.96 &   32.3 &   6.8 &   5 &  --  &  1.7 &  0.2 & y\\
13 28 14.08 &  33 34 12.6 & 22.34 & 21.70 &   38.0 &   9.7 &   4 &  --  &  2.9 &  1.3 & n\\
13 27 38.84 &  33 37 06.5 & 20.78 & 18.21 &  -55.8 &   0.8 &  51 &  --  &  3.0 &  2.9 & n\\
13 27 44.25 &  33 38 30.6 & 20.01 & 18.94 & -172.0 &   1.8 &  37 &  --  &  3.9 &  0.0 & n\\
13 27 53.03 &  33 36 00.2 & 20.62 & 18.24 &  -43.0 &   1.0 &  44 &  --  &  2.9 &  2.1 & n\\
13 28 13.43 &  33 35 40.7 & 20.52 & 19.67 & -103.9 &   1.4 &  26 &  --  &  4.0 &  0.5 & n\\
13 28 19.25 &  33 35 00.1 & 20.18 & 18.16 &   -4.2 &   1.1 &  55 &  --  &  3.6 &  1.9 & n\\
13 28 14.67 &  33 35 41.0 & 21.46 & 19.41 &   -5.7 &   1.6 &  28 &  --  &  4.0 &  1.5 & n\\
13 28 06.29 &  33 35 59.9 & 19.65 & 19.21 & -193.9 &   1.2 &  28 &  --  &  2.3 &  0.6 & n\\
13 28 12.01 &  33 37 12.6 & 21.15 & 18.50 &  -32.5 &   1.2 &  43 &  --  &  2.4 &  2.4 & n\\
13 27 29.49 &  33 35 34.4 & 22.92 & 19.80 &    2.9 &   2.3 &  28 &  --  &  1.8 &  4.0 & n\\
13 27 24.40 &  33 34 45.9 & 20.33 & 18.54 &   38.8 &   1.2 &  49 &  --  &  3.6 &  1.3 & n\\
13 27 25.64 &  33 34 01.9 & 20.30 & 18.62 &   11.2 &   3.8 &  54 &  --  &  3.8 &  1.4 & n\\
13 28 28.99 &  33 34 34.0 & 21.56 & 19.89 &  -70.6 &   1.9 &  24 &  --  &  3.2 &  0.6 & n\\
13 27 42.31 &  33 34 44.6 & 18.21 & 15.82 &  -27.6 &   1.8 &  53 &  --  &  3.2 &  2.7 & n\\
13 27 19.95 &  33 32 08.0 & 16.82 & 16.34 &  -52.1 &   3.4 &  77 &  --  &  4.1 &  0.5 & n\\
13 28 33.33 &  33 31 02.7 & 19.06 & 17.32 &  -33.7 &   1.3 &  71 &  --  &  4.6 &  1.6 & n\\
\hline
\end{tabular}
\end{center}
\end{table*}

\begin{table*}
\begin{center}
\addtocounter{table}{-1}
\caption{\emph{continued}}
\label{tableSat}
\begin{tabular}{llccccccccc}
\hline
$\alpha$ (J2000) & $\delta$ (J2000) & g & i & $v_r$ ($\kms$) & $v_{err}$ ($\kms$) & $S/N$ & $\FeH$  &
$\Sigma\textrm{Ca}$ & $\Sigma\textrm{Na}$ & member?\\\hline
13 28 19.72 &  33 29 58.0 & 18.82 & 16.58 &  -15.6 &   0.9 &  66 &  --  &  3.6 &  2.3 & n\\
13 27 18.23 &  33 29 50.5 & 18.12 & 16.50 &   -9.5 &   1.5 & 111 &  --  &  3.3 &  1.8 & n\\
13 27 28.41 &  33 29 46.6 & 21.39 & 20.52 &   41.3 &   6.8 &   5 &  --  &  2.7 &  0.3 & y\\
13 27 34.95 &  33 34 30.6 & 21.80 & 20.84 &   27.7 &   3.7 &   7 &  --  & 22.0 &  0.0 & y\\
13 27 30.36 &  33 32 04.4 & 21.45 & 20.56 &   13.8 &   2.2 &  13 & -1.9 &  2.6 &  0.0 & y\\
13 27 24.86 &  33 31 25.0 & 21.29 & 20.26 &   40.9 &   1.5 &  15 & -2.0 &  2.5 &  0.3 & y\\
13 27 24.16 &  33 31 34.3 & 21.76 & 20.92 &   31.7 &   2.6 &   9 &  --  &  2.6 &  0.1 & y\\
13 28 08.49 &  33 33 29.4 & 21.49 & 20.63 &   31.0 &   4.1 &   8 &  --  &  2.2 &  0.0 & y\\
13 27 27.05 &  33 30 30.7 & 20.76 & 19.72 &   45.0 &   1.1 &  22 & -2.2 &  2.4 &  0.3 & y\\
13 27 39.32 &  33 32 05.0 & 21.70 & 20.74 &   20.8 &   2.4 &  11 & -1.7 &  2.8 &  0.1 & y\\
13 28 12.83 &  33 30 44.7 & 20.38 & 19.15 &   21.3 &   0.9 &  31 & -1.9 &  3.3 &  0.3 & y\\
13 28 16.23 &  33 34 09.9 & 20.66 & 19.45 &   22.0 &   1.5 &  30 & -2.0 &  3.0 &  0.0 & y\\
13 28 26.82 &  33 30 59.4 & 21.37 & 20.46 &   20.0 &   2.6 &  11 & -1.8 &  2.9 &  0.2 & y\\
13 28 28.78 &  33 32 04.3 & 21.22 & 20.06 &    6.7 &   2.1 &  22 & -1.5 &  3.8 &  0.3 & y\\
13 28 11.32 &  33 32 59.2 & 21.33 & 20.23 &   22.7 &   1.9 &  12 & -1.9 &  2.7 &  0.6 & y\\
13 28 17.15 &  33 33 42.5 & 21.35 & 20.36 &   31.1 &   2.6 &  16 & -1.9 &  2.7 &  0.0 & y\\
13 28 16.61 &  33 31 00.2 & 21.56 & 20.59 &  -60.0 &   3.3 &  15 &  --  &  3.3 &  0.2 & n\\
13 28 18.51 &  33 31 42.6 & 21.17 & 20.17 &    3.5 &   1.9 &  17 & -2.0 &  2.5 &  0.3 & y\\
13 28 11.71 &  33 34 07.8 & 21.82 & 20.84 &   21.2 &   5.4 &   7 &  --  &  3.4 &  0.0 & y\\
13 27 38.76 &  33 32 55.3 & 20.52 & 19.32 &   16.8 &   1.0 &  33 & -2.3 &  2.3 &  0.5 & y\\
13 28 01.41 &  33 30 00.1 & 20.30 & 19.11 &   19.2 &   0.8 &  38 & -2.4 &  2.3 &  0.5 & y\\
13 27 55.76 &  33 32 42.2 & 20.30 & 18.96 &   21.9 &   0.8 &  39 & -1.8 &  3.5 &  0.5 & y\\
13 27 31.21 &  33 29 59.2 & 19.95 & 18.38 &   23.1 &   0.8 &  52 & -1.8 &  4.0 &  0.6 & y\\
13 27 33.80 &  33 32 57.3 & 21.22 & 20.12 &   27.2 &   1.8 &  15 & -2.0 &  2.6 &  0.2 & y\\
13 28 10.07 &  33 33 41.6 & 20.68 & 19.66 &   24.1 &   1.1 &  25 & -2.1 &  2.7 &  0.0 & y\\
13 28 07.21 &  33 30 33.0 & 20.58 & 19.50 &   45.6 &   1.0 &  28 & -2.4 &  2.0 &  0.1 & y\\
13 28 08.19 &  33 31 06.3 & 20.06 & 18.73 &   44.5 &   0.7 &  42 & -2.1 &  3.1 &  0.6 & y\\
13 27 59.22 &  33 33 06.7 & 21.43 & 20.33 &   29.8 &   2.7 &  10 &  --  &  3.0 &  0.0 & y\\
13 27 47.02 &  33 32 54.0 & 22.52 & 21.68 &   43.5 &   3.2 &   5 &  --  &  3.2 &  0.5 & y\\
13 27 42.82 &  33 30 07.0 & 22.40 & 21.46 &   26.7 &   3.3 &   5 &  --  &  2.3 &  0.5 & y\\
13 27 33.04 &  33 32 45.8 & 22.13 & 21.14 &   22.3 &   4.7 &   6 &  --  &  1.5 &  0.1 & y\\
13 27 49.23 &  33 31 32.1 & 21.86 & 21.07 &   50.2 &   7.3 &   9 &  --  &  1.3 &  0.2 & y\\
13 27 43.86 &  33 31 00.0 & 22.00 & 21.26 &   18.1 &   5.0 &   9 &  --  &  1.6 &  0.4 & y\\
13 27 54.14 &  33 30 20.8 & 22.04 & 21.12 &   33.3 &   6.7 &   8 &  --  &  2.8 &  0.3 & y\\
13 27 59.50 &  33 32 06.1 & 22.43 & 21.99 &   60.2 &   7.8 &   3 &  --  & 19.6 &  1.2 & n\\
13 27 23.82 &  33 31 23.8 & 22.37 & 21.57 &  -94.1 &   6.5 &   4 &  --  &  5.2 &  0.2 & n\\
13 28 02.45 &  33 31 09.2 & 21.91 & 21.03 &   27.5 &   2.7 &   9 &  --  &  2.9 &  0.4 & y\\
13 28 03.24 &  33 30 06.4 & 22.32 & 21.86 &  233.6 &   5.3 &   4 &  --  &  3.6 &  0.1 & n\\
13 27 23.12 &  33 32 45.7 & 22.34 & 21.41 &  196.6 &   3.0 &   4 &  --  & 10.5 &  0.7 & n\\
13 28 08.96 &  33 33 08.5 & 22.38 & 21.84 &   39.2 &   4.7 &   4 &  --  &  2.2 &  0.0 & y\\
13 27 48.31 &  33 30 46.1 & 22.08 & 21.71 &   34.5 &  11.8 &   5 &  --  &  2.4 &  0.0 & y\\
13 27 50.23 &  33 29 57.2 & 22.41 & 21.59 &   31.0 &   3.9 &   5 &  --  &  1.9 &  0.5 & y\\
13 27 51.73 &  33 30 35.8 & 22.40 & 21.93 &   40.7 &   9.7 &   4 &  --  &  3.8 &  0.7 & y\\
13 27 52.99 &  33 30 16.7 & 22.09 & 21.40 &   25.1 &   4.7 &   8 &  --  &  1.9 &  0.1 & y\\
13 28 03.54 &  33 29 54.1 & 22.34 & 21.64 &   20.1 &   6.4 &   5 &  --  &  2.5 &  0.4 & y\\
13 28 04.59 &  33 31 00.4 & 21.87 & 21.49 &   48.5 &   8.2 &   6 &  --  & 51.6 &  0.1 & y\\
13 28 12.11 &  33 33 14.9 & 22.41 & 21.58 &   20.6 &   8.5 &   4 &  --  &  4.4 &  0.0 & y\\
13 28 19.11 &  33 31 23.8 & 22.54 & 21.81 &   30.3 &   6.8 &   4 &  --  &  4.9 &  0.2 & y\\
13 28 13.25 &  33 29 49.9 & 21.93 & 21.49 &   36.0 &   6.8 &   7 &  --  &  2.6 &  0.1 & y\\
13 28 23.11 &  33 33 01.0 & 22.54 & 21.85 &   26.0 &   8.0 &   6 &  --  &  4.2 &  0.5 & y\\
13 28 14.66 &  33 33 40.6 & 21.98 & 21.02 &   24.2 &   3.0 &   7 &  --  &  2.8 &  0.0 & y\\
13 28 15.56 &  33 34 12.3 & 22.31 & 21.63 & -129.0 &  14.6 &   5 &  --  &  2.5 &  0.0 & n\\
13 27 29.58 &  33 29 45.0 & 22.04 & 19.79 &   27.6 &   1.7 &  26 &  --  &  3.6 &  1.9 & n\\
13 27 27.69 &  33 29 41.8 & 22.06 & 19.78 &   45.3 &   3.3 &  22 &  --  &  3.2 &  2.1 & n\\
13 27 25.64 &  33 34 01.9 & 20.30 & 18.62 &   11.5 &   1.1 &  48 &  --  &  4.2 &  0.0 & n\\
13 28 14.98 &  33 32 24.6 & 21.46 & 19.42 &  -32.6 &   1.8 &  29 &  --  &  3.5 &  1.5 & n\\
13 28 26.07 &  33 32 20.7 & 21.67 & 19.44 &  -11.5 &   1.4 &  30 &  --  &  3.2 &  2.2 & n\\
13 27 42.31 &  33 33 23.6 & 22.80 & 19.94 &  -24.2 &   1.6 &  26 &  --  &  2.2 &  0.0 & n\\
13 27 51.42 &  33 30 53.2 & 17.98 & 15.75 &  -45.4 &   1.1 &  57 &  --  &  3.3 &  2.1 & n\\
13 28 24.42 &  33 32 50.7 & 20.17 & 19.84 &  -36.4 &   2.0 &  25 &  --  &  1.9 &  0.2 & n\\
13 28 26.50 &  33 30 53.4 & 22.34 & 19.91 &    3.7 &   1.9 &  22 &  --  &  3.0 &  2.8 & n\\
13 28 27.13 &  33 33 09.5 & 20.24 & 18.60 &   31.3 &   2.1 &  44 &  --  &  4.0 &  1.5 & n\\
13 28 07.66 &  33 30 13.8 & 19.88 & 19.46 &  128.8 &   1.6 &  26 &  --  &  2.3 &  0.4 & n\\
\hline
\end{tabular}
\end{center}
\end{table*}

\begin{table*}
\begin{center}
\caption{Derived parameters for stars in the Ursa Major~I sample.}
\label{tableSat}
\begin{tabular}{llccccccccc}
\hline
$\alpha$ (J2000) & $\delta$ (J2000) & g & i & $v_r$ ($\kms$) & $v_{err}$ ($\kms$) & $S/N$ & $\FeH$  &
$\Sigma\textrm{Ca}$ & $\Sigma\textrm{Na}$ & member?\\\hline
10 34 11.53 &  51 54 22.1 & 15.74 & 15.14 &   51.9 &   1.7 &  83 &  --  &  4.0 &  0.5 & n\\
10 35 33.79 &  51 56 46.6 & 16.19 & 15.59 &   -9.9 &   1.6 &  58 &  --  &  4.6 &  0.8 & n\\
10 34 17.26 &  51 57 33.6 & 20.16 & 19.25 &  -59.2 &   1.1 &  28 & -2.4 &  1.1 &  0.1 & y\\
10 34 22.13 &  51 56 10.6 & 20.56 & 19.78 &  -39.6 &   2.2 &  20 & -1.5 &  2.9 &  0.2 & y\\
10 34 18.21 &  51 56 33.8 & 21.43 & 20.75 &  -76.4 &   4.9 &   9 &  --  &  1.2 &  0.2 & y\\
10 34 31.15 &  51 57 01.1 & 18.89 & 17.75 &  -57.2 &   0.7 &  26 & -2.5 &  1.7 &  0.2 & y\\
10 34 58.15 &  51 55 23.8 & 20.56 & 19.72 &  -59.6 &   5.6 &   9 &  --  &  1.0 &  0.2 & y\\
10 35 03.55 &  51 56 20.1 & 20.91 & 20.12 &  -70.0 &   2.4 &  12 & -2.2 &  1.1 &  0.2 & y\\
10 35 17.58 &  51 55 33.9 & 19.47 & 18.37 &  -70.3 &   1.2 &  29 &  --  &  5.3 &  0.1 & n\\
10 35 28.87 &  51 57 01.5 & 18.63 & 17.44 &  -57.2 &   0.6 &  25 & -2.4 &  2.2 &  0.0 & y\\
10 34 33.82 &  51 55 52.2 & 21.41 & 20.86 & -137.5 &   2.5 &   8 &  --  &  2.2 &  0.3 & n\\
10 35 31.46 &  51 57 11.0 & 18.67 & 17.48 &   49.9 &   1.9 &  41 &  --  &  3.2 &  0.7 & n\\
10 35 42.86 &  51 56 17.7 & 20.06 & 19.16 &  -70.1 &   1.3 &  21 & -2.0 &  2.1 &  0.1 & y\\
10 35 23.72 &  51 57 16.1 & 21.77 & 20.99 & -482.6 &  14.8 &   3 &  --  &  1.9 &  0.7 & n\\
10 34 26.48 &  51 55 32.3 & 22.18 & 21.56 &  -69.0 &   4.3 &   6 &  --  &  1.5 &  0.4 & y\\
10 34 29.72 &  51 54 37.6 & 22.03 & 21.42 &  -42.2 &  11.1 &   5 &  --  &  1.2 &  0.0 & y\\
10 34 46.10 &  51 56 06.0 & 22.07 & 21.39 & -100.8 &  10.8 &   4 &  --  &  2.9 &  0.5 & n\\
10 35 09.99 &  51 57 30.9 & 21.74 & 21.04 &  -74.5 &   3.6 &   6 &  --  &  1.2 &  0.0 & y\\
10 34 42.93 &  51 56 15.2 & 21.86 & 21.08 &  -58.0 &   2.9 &   6 &  --  &  1.2 &  0.3 & y\\
10 34 02.55 &  51 54 17.6 & 19.24 & 18.49 &  -86.2 &   5.1 &  26 &  --  &  5.5 &  0.5 & n\\
10 34 30.44 &  51 54 23.5 & 20.90 & 17.98 &  -13.4 &   1.9 &  19 &  --  &  1.3 &  3.3 & n\\
10 33 59.76 &  51 54 39.8 & 21.08 & 19.38 & -147.1 &   3.7 &  52 &  --  &  0.7 &  0.2 & n\\
10 34 23.33 &  51 55 57.0 & 20.07 & 17.91 &   -8.9 &   1.5 &  25 &  --  &  2.7 &  1.7 & n\\
10 34 16.12 &  51 57 17.5 & 22.64 & 19.83 &  -17.7 &   2.2 &  23 &  --  &  2.3 &  2.5 & n\\
10 35 04.73 &  51 54 22.8 & 16.60 & 14.38 &  -18.1 &   1.1 &  78 &  --  &  3.8 &  2.3 & n\\
10 34 37.01 &  51 54 39.1 & 20.02 & 19.28 & -193.5 &   2.0 &  25 &  --  &  4.2 &  0.9 & n\\
10 34 59.36 &  51 54 45.1 & 19.01 & 18.39 &  -43.2 &   1.1 &  28 &  --  &  1.9 &  0.4 & n\\
10 35 02.33 &  51 55 47.8 & 20.56 & 18.44 &   14.3 &   1.5 &  24 &  --  &  3.1 &  1.7 & n\\
10 34 53.68 &  51 54 54.4 & 20.89 & 19.77 &  -63.5 &   2.2 &  12 &  --  &  1.4 &  0.5 & n\\
10 35 10.81 &  51 55 41.9 & 19.12 & 18.69 &  -75.0 &   1.9 &  30 &  --  &  2.2 &  0.2 & n\\
10 34 54.62 &  51 54 51.7 & 18.57 & 17.90 &  -33.1 &   1.5 &  24 &  --  &  2.9 &  0.1 & n\\
10 34 44.31 &  51 55 26.3 & 21.79 & 19.99 & -176.8 &   2.5 &  18 &  --  &  3.9 &  0.8 & n\\
10 34 40.95 &  51 57 09.6 & 22.22 & 19.15 &   -9.4 &   1.2 &  29 &  --  &  1.7 &  2.8 & n\\
10 35 22.53 &  51 56 17.6 & 21.59 & 19.09 &  -36.8 &   1.6 &  27 &  --  &  2.5 &  2.2 & n\\
10 34 27.92 &  51 58 47.1 & 21.55 & 19.22 &  -18.0 &   1.5 &  30 &  --  &  2.8 &  0.0 & n\\
10 34 00.14 &  51 51 59.3 & 19.05 & 16.69 &   12.3 &   1.3 &  52 &  --  &  3.3 &  2.6 & n\\
10 34 14.38 &  51 51 07.6 & 18.29 & 17.33 &   64.3 &   1.2 &  39 &  --  &  4.0 &  1.0 & n\\
10 35 25.86 &  51 51 16.2 & 19.31 & 17.14 &   22.4 &   1.0 &  79 &  --  &  4.0 &  2.6 & n\\
10 35 23.85 &  51 53 28.0 & 19.01 & 16.94 &    0.6 &   1.8 &  74 &  --  &  4.2 &  2.6 & n\\
10 34 02.73 &  51 53 25.1 & 20.90 & 20.27 &  153.8 &   6.7 &   9 &  --  &  3.1 &  0.3 & n\\
10 34 48.11 &  51 50 50.9 & 20.93 & 20.13 &  -56.0 &   1.8 &  12 & -2.0 &  1.6 &  0.1 & y\\
10 34 50.73 &  51 52 40.2 & 21.10 & 20.29 &  -45.5 &   3.1 &  12 & -1.4 &  2.8 &  0.2 & y\\
10 35 06.91 &  51 52 25.1 & 19.61 & 18.62 &  -42.9 &   2.5 &  36 &  --  &  4.0 &  1.1 & n\\
10 35 11.49 &  51 51 33.1 & 19.77 & 18.92 &  -55.3 &   1.2 &  32 & -2.3 &  1.5 &  0.3 & y\\
10 35 42.36 &  51 52 16.5 & 20.11 & 19.24 &   12.0 &   2.8 &  24 &  --  &  3.2 &  0.5 & n\\
10 35 36.70 &  51 53 18.4 & 21.18 & 20.46 &  -55.7 &   2.6 &  11 & -2.0 &  1.3 &  0.2 & y\\
10 34 08.46 &  51 50 48.7 & 22.46 & 21.98 & -130.8 &  12.7 &   2 &  --  &  1.6 &  1.0 & n\\
10 34 06.46 &  51 52 32.8 & 22.00 & 21.27 & -405.0 &   5.7 &   5 &  --  &  1.4 &  0.6 & n\\
10 34 16.41 &  51 55 28.6 & 22.33 & 21.71 &   96.9 &   6.3 &   5 &  --  &  1.4 &  0.0 & n\\
10 34 38.19 &  51 53 39.2 & 21.70 & 21.18 & -195.3 &   9.9 &   5 &  --  &  2.8 &  1.5 & n\\
10 35 19.13 &  51 53 59.6 & 22.29 & 21.67 &  -63.3 &   7.9 &   4 &  --  & 26.7 &  0.4 & y\\
10 34 03.70 &  51 50 50.8 & 21.04 & 18.72 &   89.5 &   2.1 &  36 &  --  &  3.3 &  1.8 & n\\
10 34 09.85 &  51 52 01.5 & 20.94 & 18.87 &  -25.0 &   1.7 &  29 &  --  &  3.3 &  1.8 & n\\
10 34 19.48 &  51 53 06.8 & 19.56 & 17.55 &  -20.4 &   1.1 &  40 &  --  &  4.3 &  2.1 & n\\
10 34 34.92 &  51 51 22.7 & 20.62 & 18.66 &  -26.3 &   1.8 &  38 &  --  &  3.5 &  1.8 & n\\
10 34 30.44 &  51 54 23.5 & 20.90 & 17.98 &   -8.3 &   1.7 &  42 &  --  &  1.6 &  0.0 & n\\
10 34 11.32 &  51 51 49.4 & 21.57 & 18.84 &   55.9 &   2.2 &  40 &  --  &  1.8 &  3.2 & n\\
\hline
\end{tabular}
\end{center}
\end{table*}

\begin{table*}
\begin{center}
\addtocounter{table}{-1}
\caption{\emph{continued}}
\label{tableSat}
\begin{tabular}{llccccccccc}
\hline
$\alpha$ (J2000) & $\delta$ (J2000) & g & i & $v_r$ ($\kms$) & $v_{err}$ ($\kms$) & $S/N$ & $\FeH$  &
$\Sigma\textrm{Ca}$ & $\Sigma\textrm{Na}$ & member?\\\hline
10 34 23.10 &  51 52 43.0 & 19.03 & 18.68 &  107.3 &   1.9 &  36 &  --  &  1.9 &  0.7 & n\\
10 34 37.01 &  51 54 39.1 & 20.02 & 19.28 & -186.8 &   1.8 &  26 &  --  &  4.1 &  0.0 & n\\
10 34 47.55 &  51 52 30.2 & 19.94 & 17.61 &  -28.3 &   1.2 &  51 &  --  &  3.1 &  2.7 & n\\
10 35 10.01 &  51 51 17.9 & 19.54 & 19.12 &  -38.4 &   1.4 &  29 &  --  &  1.8 &  0.4 & n\\
10 35 13.69 &  51 50 57.7 & 20.56 & 17.78 &   51.0 &   1.1 &  46 &  --  &  2.2 &  3.1 & n\\
10 35 02.34 &  51 53 09.1 & 21.67 & 18.86 &   48.1 &   4.5 &  25 &  --  &  2.1 &  3.0 & n\\
10 35 12.06 &  51 53 35.6 & 22.28 & 19.93 &  -21.5 &   2.7 &  17 &  --  &  2.8 &  1.7 & n\\
10 35 16.46 &  51 52 15.5 & 21.90 & 19.58 &  -95.5 &   4.6 &  25 &  --  &  3.4 &  2.0 & n\\
10 35 39.05 &  51 50 49.9 & 21.61 & 19.01 &   19.5 &   2.0 &  33 &  --  &  2.8 &  2.8 & n\\
10 35 25.33 &  51 52 40.7 & 19.56 & 17.90 &   42.4 &   0.9 &  49 &  --  &  4.1 &  1.7 & n\\
10 34 36.19 &  51 53 29.3 & 18.13 & 17.55 &  -19.1 &   1.0 &  57 &  --  &  3.8 &  1.0 & n\\
\hline
\end{tabular}
\end{center}
\end{table*}

\begin{table*}
\begin{center}
\caption{Derived parameters for stars in the Ursa Major~II sample.}
\label{tableSat}
\begin{tabular}{llccccccccc}
\hline
$\alpha$ (J2000) & $\delta$ (J2000) & g & i & $v_r$ ($\kms$) & $v_{err}$ ($\kms$) & $S/N$ & $\FeH$  &
$\Sigma\textrm{Ca}$ & $\Sigma\textrm{Na}$ & member?\\\hline
08 50 46.29 &  63 04 48.6 & 19.46 & 17.39 &   -5.4 &   1.6 &  42 &  --  &  3.6 &  1.8 & n\\
08 50 19.61 &  63 05 52.0 & 19.41 & 16.92 &    5.2 &   1.0 &  70 &  --  &  3.5 &  2.7 & n\\
08 51 54.46 &  63 05 57.2 & 18.08 & 16.62 &    3.9 &   1.1 &  53 &  --  &  4.9 &  1.5 & n\\
08 52 17.54 &  63 05 14.8 & 18.60 & 16.05 &    2.4 &   1.4 &  49 &  --  &  2.7 &  2.8 & n\\
08 49 57.80 &  63 03 56.1 & 17.70 & 16.57 &  -50.0 &   1.0 &  76 &  --  &  4.2 &  1.0 & n\\
08 50 21.25 &  63 04 08.7 & 20.68 & 19.96 &  -15.3 &   9.9 &  11 &  --  &  3.2 &  0.3 & n\\
08 49 58.95 &  63 04 22.9 & 21.11 & 20.56 &  -99.2 &  13.7 &   5 &  --  &  4.0 &  0.3 & y\\
08 50 58.80 &  63 05 54.4 & 18.50 & 17.97 &  -56.5 &   1.0 &  42 &  --  &  2.4 &  0.7 & n\\
08 51 03.18 &  63 05 45.2 & 16.54 & 15.63 &   84.8 &   1.1 &  59 &  --  &  4.4 &  1.0 & n\\
08 51 14.04 &  63 06 27.8 & 18.77 & 17.77 &  -46.5 &   1.3 &  42 &  --  &  4.0 &  1.1 & n\\
08 50 12.87 &  63 05 58.8 & 21.06 & 20.26 & -148.4 &   3.9 &  10 &  --  &  0.9 &  0.9 & n\\
08 50 18.29 &  63 05 37.1 & 21.03 & 20.28 &  -63.8 &   2.9 &   9 &  --  &  3.0 &  0.1 & n\\
08 50 41.90 &  63 06 59.6 & 21.50 & 20.86 & -125.4 &   3.4 &   6 &  --  &  2.6 &  0.8 & y\\
08 51 17.07 &  63 03 47.3 & 20.15 & 19.22 & -110.6 &   1.7 &  16 & -1.8 &  1.2 &  0.8 & y\\
08 52 08.97 &  63 06 51.3 & 21.47 & 20.85 &  -92.2 &   7.4 &   6 &  --  &  2.0 &  0.4 & y\\
08 51 47.21 &  63 04 25.8 & 16.59 & 15.68 &  -46.7 &   1.1 &  44 &  --  &  4.3 &  0.9 & n\\
08 51 27.18 &  63 07 03.1 & 21.19 & 20.53 &  -57.7 &   4.6 &   8 &  --  &  2.7 &  0.0 & n\\
08 52 09.72 &  63 04 41.6 & 21.43 & 20.95 & -151.5 &   3.6 &   6 &  --  &  2.0 &  1.5 & n\\
08 52 19.57 &  63 03 46.2 & 21.27 & 20.51 & -113.8 &  13.1 &   8 &  --  &  1.0 &  0.2 & y\\
08 51 48.45 &  63 06 44.7 & 21.59 & 20.95 & -112.4 &   3.2 &   5 &  --  &  1.9 &  0.7 & y\\
08 52 16.52 &  63 04 06.2 & 20.44 & 19.59 & -123.2 &   2.5 &  16 & -1.8 &  0.8 &  0.3 & y\\
08 52 12.28 &  63 05 35.2 & 19.66 & 18.80 & -116.3 &   1.1 &  26 & -1.8 &  1.4 &  0.8 & y\\
08 51 55.69 &  63 08 21.8 & 20.93 & 20.17 &  126.3 &   6.5 &  12 &  --  &  4.2 &  0.0 & n\\
08 49 59.59 &  63 03 48.3 & 21.92 & 21.40 & -587.0 &  13.4 &   3 &  --  &  1.6 &  2.1 & n\\
08 50 02.28 &  63 05 57.6 & 22.33 & 21.60 &  501.8 &   9.5 &   2 &  --  &  3.3 &  0.5 & n\\
08 50 08.12 &  63 06 46.3 & 21.88 & 21.20 & -125.6 &   3.9 &   4 &  --  &  0.2 &  0.7 & y\\
08 50 04.44 &  63 06 14.6 & 22.36 & 21.75 &  247.5 &  10.6 &   3 &  --  &  3.9 &  0.9 & n\\
08 50 48.69 &  63 07 33.0 & 21.64 & 21.07 & -109.2 &   5.2 &   4 &  --  &  1.2 &  0.0 & y\\
08 51 16.15 &  63 06 19.5 & 21.78 & 21.14 &  -79.4 &   3.5 &   5 &  --  &  3.0 &  0.8 & n\\
08 51 19.57 &  63 03 55.9 & 21.90 & 21.39 & -286.7 &  14.2 &   3 &  --  &  2.2 &  0.8 & n\\
08 51 29.23 &  63 04 15.7 & 21.73 & 21.12 & -103.3 &   5.2 &   5 &  --  &  2.4 &  1.1 & y\\
08 51 52.02 &  63 05 08.6 & 21.61 & 21.21 &  203.9 &   6.5 &   5 &  --  &  0.9 &  0.5 & n\\
08 51 42.27 &  63 04 40.1 & 21.98 & 21.55 & -396.0 &   6.2 &   3 &  --  &  0.1 &  1.6 & n\\
08 50 24.93 &  63 03 56.9 & 21.49 & 19.22 &  -56.2 &   1.8 &  23 &  --  &  3.7 &  1.8 & n\\
08 50 27.01 &  63 04 34.1 & 20.52 & 17.74 &   -8.5 &   1.0 &  46 &  --  &  2.1 &  3.3 & n\\
08 50 54.74 &  63 03 46.5 & 18.01 & 15.95 &  -24.9 &   1.1 &  56 &  --  &  4.6 &  2.1 & n\\
08 51 02.04 &  63 05 28.6 & 20.41 & 17.81 &  -32.6 &   0.9 &  44 &  --  &  3.0 &  2.6 & n\\
08 50 56.08 &  63 06 17.9 & 19.42 & 17.64 &   38.2 &   1.3 &  49 &  --  &  4.5 &  2.1 & n\\
08 50 20.52 &  63 05 13.3 & 17.35 & 15.83 &  -70.0 &   1.2 &  67 &  --  &  5.3 &  1.5 & n\\
08 50 13.55 &  63 05 35.9 & 21.15 & 19.96 & -116.4 &   7.2 &   9 &  --  &  4.5 &  0.6 & n\\
08 50 16.55 &  63 06 58.3 & 21.60 & 19.36 &  -17.5 &   2.1 &  20 &  --  &  3.8 &  1.9 & n\\
08 51 05.36 &  63 05 21.6 & 21.11 & 18.51 &   17.2 &   1.3 &  36 &  --  &  2.8 &  3.0 & n\\
08 50 45.24 &  63 06 06.0 & 21.30 & 18.46 &   -1.2 &   1.3 &  34 &  --  &  2.5 &  3.5 & n\\
08 50 09.75 &  63 05 41.8 & 22.58 & 19.73 &  -23.3 &   2.9 &  19 &  --  &  1.8 &  3.4 & n\\
08 50 50.63 &  63 07 08.5 & 22.34 & 19.47 &   -3.6 &   1.8 &  22 &  --  &  1.8 &  0.0 & n\\
08 51 44.48 &  63 03 45.3 & 19.23 & 17.69 &  -60.1 &   1.0 &  52 &  --  &  4.6 &  1.2 & n\\
08 51 21.43 &  63 05 55.7 & 22.16 & 19.84 &  -76.8 &   8.4 &  12 &  --  &  3.9 &  1.3 & n\\
08 52 05.07 &  63 03 38.0 & 19.83 & 17.55 &   72.7 &   1.0 &  54 &  --  &  3.8 &  1.4 & n\\
08 51 49.67 &  63 03 44.8 & 20.14 & 20.00 & -126.1 &   7.8 &  13 &  --  &  1.9 &  0.6 & n\\
08 52 02.99 &  63 05 47.1 & 21.77 & 19.09 &  104.5 &   6.2 &  28 &  --  &  1.4 &  2.7 & n\\
08 51 41.72 &  63 06 21.0 & 21.78 & 18.83 &  -40.3 &   1.8 &  24 &  --  &  2.1 &  3.2 & n\\
08 52 15.16 &  63 04 16.3 & 21.79 & 19.14 &   12.8 &   1.5 &  25 &  --  &  2.9 &  2.5 & n\\
08 51 57.96 &  63 08 17.5 & 22.21 & 19.84 &  -49.2 &   4.3 &  19 &  --  &  3.2 &  0.0 & n\\
08 52 13.09 &  63 06 55.9 & 22.33 & 19.30 &   38.1 &   2.7 &  21 &  --  &  2.2 &  3.5 & n\\
\hline
\end{tabular}
\end{center}
\end{table*}

\begin{table*}
\begin{center}
\caption{Derived parameters for stars in the Willman~1 sample.}
\label{tableSat}
\begin{tabular}{llccccccccc}
\hline
$\alpha$ (J2000) & $\delta$ (J2000) & g & i & $v_r$ ($\kms$) & $v_{err}$ ($\kms$) & $S/N$ & $\FeH$  &
$\Sigma\textrm{Ca}$ & $\Sigma\textrm{Na}$ & member?\\\hline
10 50 15.74 &  51 02 22.3 & 15.96 & 15.08 &  -51.5 &   1.1 &  88 &  --  &  5.2 &  0.9 & n\\
10 48 39.51 &  51 04 35.7 & 19.85 & 17.41 &    8.7 &   1.3 &  66 &  --  &  3.5 &  2.5 & n\\
10 48 46.16 &  51 02 12.2 & 19.27 & 17.42 &   55.4 &   1.1 &  65 &  --  &  3.9 &  1.5 & n\\
10 49 56.99 &  51 05 49.5 & 19.36 & 16.66 &   15.7 &   1.2 &  65 &  --  &  2.2 &  0.0 & n\\
10 49 12.41 &  51 05 44.2 & 18.90 & 18.01 &  -10.2 &   1.2 &  48 & -0.8 &  4.6 &  0.0 & y\\
10 49 17.43 &  51 03 25.9 & 20.55 & 19.90 &  -16.7 &   3.9 &  15 & -1.7 &  1.4 &  0.1 & y\\
10 49 16.76 &  51 04 03.6 & 21.39 & 20.87 &  -14.0 &   4.9 &   7 &  --  &  2.9 &  0.4 & y\\
10 49 21.16 &  51 03 30.2 & 21.20 & 20.67 &  -22.3 &  14.7 &   8 &  --  &  1.8 &  0.2 & y\\
10 49 08.09 &  51 02 27.0 & 21.04 & 20.32 &   -8.3 &   2.8 &  13 & -1.1 &  2.5 &  0.1 & y\\
10 49 15.98 &  51 02 26.5 & 21.05 & 20.33 &  -12.2 &   4.8 &  12 & -1.1 &  2.3 &  0.3 & y\\
10 48 58.13 &  51 02 53.9 & 21.36 & 20.75 &  -13.4 &   3.8 &   9 &  --  &  1.7 &  0.0 & y\\
10 49 27.86 &  51 03 46.3 & 20.74 & 20.21 &  -11.7 &   2.8 &  15 & -1.6 &  1.4 &  0.1 & y\\
10 49 42.89 &  51 04 22.8 & 18.87 & 17.96 &  -13.2 &   1.0 &  49 &  --  &  4.5 &  1.1 & n\\
10 49 52.54 &  51 03 42.5 & 18.75 & 17.89 &  -22.0 &   0.6 &  55 & -2.1 &  1.6 &  0.4 & y\\
10 49 40.83 &  51 03 40.3 & 21.31 & 20.78 &   -6.9 &   7.1 &  10 & -1.4 &  1.4 &  0.2 & y\\
10 49 30.96 &  51 03 41.0 & 21.34 & 20.81 &    7.3 &  14.1 &   8 &  --  &  2.3 &  0.1 & n\\
10 49 10.13 &  51 03 00.3 & 21.32 & 21.15 &   -4.9 &   6.8 &  12 & -1.4 &  1.3 &  0.4 & y\\
10 49 24.36 &  51 02 29.3 & 21.49 & 21.38 & -442.4 &   7.8 &   4 &  --  &  2.4 &  0.2 & n\\
10 48 57.14 &  51 02 31.4 & 21.83 & 21.39 &   -2.4 &   7.9 &   6 &  --  &  1.5 &  0.1 & y\\
10 49 23.67 &  51 03 03.9 & 22.45 & 21.91 & -173.1 &   7.3 &   3 &  --  &  1.9 &  1.0 & n\\
10 49 21.62 &  51 02 45.3 & 22.21 & 21.81 &  230.6 &   3.0 &   4 &  --  &  1.4 &  0.1 & n\\
10 49 20.26 &  51 03 42.9 & 21.88 & 21.62 &   59.5 &   9.4 &   4 &  --  &  2.9 &  0.2 & n\\
10 49 25.04 &  51 02 25.2 & 21.91 & 21.47 &  208.7 &   5.2 &   4 &  --  &  2.0 &  1.0 & n\\
10 49 26.33 &  51 03 16.4 & 21.81 & 21.49 & -374.9 &   8.4 &   4 &  --  & 19.4 &  0.3 & n\\
10 49 31.41 &  51 03 02.7 & 21.51 & 21.31 &  498.4 &   7.4 &   9 &  --  &  0.7 &  0.3 & n\\
10 49 47.37 &  51 03 34.7 & 22.00 & 21.87 &   -8.5 &   8.5 &   4 &  --  &  1.9 &  0.7 & y\\
10 49 50.13 &  51 05 12.5 & 22.11 & 21.76 &  542.8 &   8.7 &   4 &  --  &  0.5 &  0.4 & n\\
10 50 02.89 &  51 02 32.2 & 21.67 & 21.09 &   -7.7 &   4.1 &   8 &  --  &  3.0 &  0.1 & y\\
10 49 39.00 &  51 04 57.4 & 22.14 & 21.85 &  -74.1 &   7.7 &   4 &  --  &  3.7 &  0.4 & n\\
10 49 34.78 &  51 04 28.5 & 21.91 & 21.61 &   12.7 &  11.2 &   5 &  --  &  1.8 &  0.4 & n\\
10 49 07.40 &  51 04 09.8 & 21.85 & 19.81 &   -2.2 &   2.0 &  22 &  --  &  3.4 &  1.9 & n\\
10 49 06.85 &  51 04 23.6 & 18.77 & 18.33 &  220.4 &   1.1 &  45 &  --  &  2.5 &  0.4 & n\\
10 49 03.88 &  51 06 40.2 & 17.69 & 15.50 &  -10.2 &   1.1 &  84 &  --  &  4.0 &  0.0 & n\\
10 48 45.14 &  51 05 36.2 & 21.02 & 18.78 &  -39.3 &   2.0 &  34 &  --  &  3.2 &  2.1 & n\\
10 48 51.58 &  51 03 09.9 & 19.77 & 19.32 &   38.6 &   1.4 &  25 &  --  &  2.2 &  0.3 & n\\
10 48 48.96 &  51 04 15.5 & 22.00 & 19.02 &  -19.8 &   1.2 &  37 &  --  &  2.1 &  4.0 & n\\
10 48 53.25 &  51 06 30.6 & 21.46 & 18.82 &  -24.0 &   1.2 &  38 &  --  &  2.6 &  0.0 & n\\
10 48 53.65 &  51 06 23.8 & 21.67 & 18.82 &  -26.3 &   1.8 &  37 &  --  &  2.3 &  0.0 & n\\
10 48 50.57 &  51 04 47.7 & 22.18 & 19.39 &  -48.9 &   2.3 &  31 &  --  &  2.1 &  2.9 & n\\
10 48 58.89 &  51 04 56.6 & 22.76 & 19.88 &   10.8 &   2.2 &  21 &  --  &  1.8 &  3.9 & n\\
10 49 19.54 &  51 04 17.3 & 22.28 & 19.71 &  -44.0 &   3.4 &  25 &  --  &  2.4 &  2.5 & n\\
10 49 33.83 &  51 03 33.8 & 20.83 & 19.53 &   29.6 &   2.1 &  23 &  --  &  4.3 &  1.2 & n\\
10 50 09.51 &  51 04 15.1 & 19.93 & 17.73 &  -94.8 &   2.2 &  46 &  --  &  3.4 &  1.8 & n\\
10 49 51.71 &  51 05 11.8 & 21.99 & 19.70 &   14.6 &   2.4 &  22 &  --  &  2.8 &  2.2 & n\\
10 49 54.59 &  51 04 16.7 & 22.42 & 19.78 &   28.1 &   1.5 &  25 &  --  &  2.8 &  2.7 & n\\
10 50 04.57 &  51 03 33.0 & 22.93 & 19.98 &  -31.5 &   5.5 &  24 &  --  &  1.2 &  3.5 & n\\
10 50 07.87 &  51 02 57.4 & 19.16 & 17.81 &  -28.5 &   1.1 &  45 &  --  &  4.8 &  1.5 & n\\
10 50 01.31 &  51 04 43.4 & 21.64 & 18.44 &   14.6 &   1.6 &  38 &  --  &  1.1 &  4.5 & n\\
\hline
\end{tabular}
\end{center}
\end{table*}

\begin{table*}
\begin{center}
\caption{Parameters used and derived for the Milky Way satellites observed and analyzed in this paper. From top to bottom are
listed their heliocentric distance, half-light radius, heliocentric systemic velocity, their systemic velocity corrected from the solar motion, their velocity dispersion, the V magnitude of the horizontal branch of the satellite that was used to derive the metallicity of their members, their median metallicity and their total mass derived through equation (3). In the case of CVnI, the parameters
of the cold metal-rich and of the hot metal-poor components found by \citet{ibata06} are listed. For the other satellites, distances and half-light radii are taken from the references cited in the text.}
\label{tableSat}
\begin{tabular}{l|cccccc}
\hline\hline
 & Boo & CVnI (cold comp.) & CVnI (hot comp.)  & UMaI & UMaII & Wil1 \\
$D$ (kpc) & $62\pm3$ & $\sim220$ & $\sim220$ & $\sim100$ & $30\pm5$ & $38\pm7$\\
$r_{hb}$ (pc) & $\sim230$ & $\sim230$ & $\sim500$ & $\sim250$ & $50-120$ & $\sim20$\\
$v_r$ ($\kms$) & $99.0\pm2.1$ & $22.5\pm0.5$ & $26.5\pm1.5$ & $-57.0\pm3.5$ & $-115\pm5$ & $-12.3\pm2.5$\\
$v_{gsr}$ ($\kms$) & $106.5\pm2.1$ & $68.3\pm0.5$ & $72.3\pm1.5$ & $-10.6\pm3.5$ & $-35\pm5$ & $32.4\pm2.5$\\
$\sigma_{vr}$ ($\kms$) & $6.5_{-1.4}^{+2.0}$ & $0.5\pm0.5$ & $13.9_{-2.5}^{+3.2}$ & $11.9_{-2.3}^{+3.5}$ (a) &
$7.4_{-2.8}^{+4.5}$ & $4.3_{-1.3}^{+2.3}$\\
$V_{HB}$ (b) & 19.5 & 22.4 & 22.4 & 20.5 & 18.1 & 18.6\\
Median $\FeH$ & $-2.1$ & $\sim-1.7$ & $\sim-2.1$ & $-2.0$ to $-2.4$ & $-1.8$ (c) & $-1.5$\\
Mass ($\msun$) & $1.3\times10^7$ & see \citet{ibata06} & see \citet{ibata06} & $4.7\times10^7$ & -- & $5\times10^5$\\
\hline
\end{tabular}
\end{center} (a) Note however that UMaI may contain a population with a dispersion consistent with $0\kms$; (b) $V_{HB}$
has been
determined from the SDSS CMDs for satellites that show such a feature (Boo, CVn, UMaI) or as $m-M-0.7$
otherwise; (c) for UMaII, it has to be noted that members were selected according to their metallicity so the median
metallicity given here is by selection similar to previous estimates.
\end{table*}

\end{document}